\documentclass[journal]{IEEEtran}
\usepackage[dvips]{graphics}
\usepackage{epsfig}
\usepackage{amsmath}

\ifCLASSINFOpdf
\else
\fi
\begin{document}
\title{A Kalman Filtering approach of improved precision
for fault diagnosis in distributed parameter systems}

\author{Gerasimos G. Rigatos,~\IEEEmembership{Member,~IEEE}
        \thanks{G. Rigatos is with the Unit of Industrial Automation, Industrial Systems Institute, 26504, Rion Patras, Greece, Email: grigat@ieee.org.}}% <-this % stops a space}


\markboth{}%
{Shell \MakeLowercase{\textit{et al.}}: Bare Demo of IEEEtran.cls for Journals}



\maketitle

\begin{abstract}
%\boldmath
The Derivative-free nonlinear Kalman Filter is proposed for state estimation and fault diagnosis in distributed parameter systems and particularly in dynamical systems described by partial differential equations of the nonlinear wave type. At a first stage, a nonlinear filtering approach for estimating the dynamics of a 1D nonlinear wave equation, from measurements provided from a small number of sensors is developed. It is shown that the numerical solution of the associated partial differential equation results into a set of nonlinear ordinary differential equations. With the application of diffeomorphism that is based on differential flatness theory it is shown that an equivalent description of the system is obtained in the linear canonical (Brunovsky) form. This transformation enables to obtain local estimates about the state vector of the system through the application of the standard Kalman Filter recursion. At a second stage, the local statistical approach to fault diagnosis is used to perform fault diagnosis for the distributed parameters system by processing with elaborated statistical tools the differences (residuals) between the output of the Kalman Filter and  the measurements obtained from the distributed parameter system. Optimal selection of the fault threshold is succeeded by using the local statistical approach to fault diagnosis. The efficiency of the proposed filtering approach for performing fault diagnosis in distributed parameters systems is confirmed through simulation experiments.
\end{abstract}

\begin{IEEEkeywords}
distributed parameter systems, differential flatness theory, derivative-free nonlinear Kalman Filtering, nonlinear wave equations, local statistical approach to fault diagnosis, fault detection, fault isolation.
\end{IEEEkeywords}

% \fi
\IEEEpeerreviewmaketitle

\section{Introduction} \label{section 1: Introduction}

\noindent Fault diagnosis in distributed parameter systems is a complicated problem that has been little explored up to now. A reason for this is that state estimation methods used for residual generation in distributed parameter systems and in infinite dimensional systems described by partial differential equations are much more complicated than state estimation methods for lumped parameter systems [\ref{WoiRud12}-\ref{HaiRam10}]. Of particular interest is state estimation of wave-type nonlinear phenomena, appearing in several engineering applications [\ref{HidBabSchuNun11}-\ref{Chau11}]. The paper treats the problem of estimation and fault diagnosis for 1D nonlinear infinite dimensional systems, which are described by Partial Differential Equations (PDEs) of the wave type.

\noindent At a first stage the dynamics of the PDE model is computed through a state estimator that processes measurements from a small number of sensors. To this end the following steps are followed. Using the method for numerical solution of the PDE through discretization the initial partial differential equation is decomposed into a set of nonlinear ordinary differential equations with respect to time [\ref{Pin91}]. Next, each one of the local models associated with the ordinary differential equations is transformed into a model of the linear canonical (Brunovsky) form through a change of coordinates (diffeomorphism) which is based on differential flatness theory. This transformation provides an extended model of the nonlinear PDE for which state estimation is possible by application of the standard Kalman Filter recursion [\ref{MouRud01}-\ref{MarTom92}]. Unlike other nonlinear estimation methods (e.g. Extended Kalman Filter) the application of the standard Kalman Filter recursion to the linearized equivalent of the nonlinear PDE system does need the computation of Jacobian matrices and partial derivatives [\ref{Rig11}-\ref{Rig12a}].

\noindent At a second stage, the paper proposes the \textit{local statistical approach to fault diagnosis} for detecting changes and faults in the distributed parameter system  [\ref{ZhaBasBen98}-\ref{BasBenZha96}]. Residuals are generated by comparing the outputs measured from the distributed parameter system against the outputs obtained from the derivative-free nonlinear Kalman Filter. The processing of these differences through the local statistical approach to fault diagnosis provides clear indications about the existence of incipient changes in the model of the monitored PDE. Fault diagnosis with the Local Statistical Approach has two significant advantages: i) it provides a credible criterion ($\chi^2$ test) to detect if faults have taken place in the distributed parameters system. This criterion is more efficient than the normalized square error and mean error tests since it employs the modeling error derivative and records the tendency for change. Thus early change detection for the filter's parameters becomes possible ii) it recognizes the parameters of the PDE model which are responsible for the deviation of the filter's estimates from the real output of the monitored dynamical system.

\noindent The structure of the paper is as follows: In Section \ref{section 2: Filtering_using_differential_flatness_theory_and_canonical_forms} nonlinear filtering using differential flatness theory and transformation of the system's dynamics into canonical forms is analyzed. A new filtering method, under the name Derivative-free distributed nonlinear Kalman Filtering, is proposed for distributed state estimation. In Section \ref{Section 3: Estimation_of_nonlinear_wave_equations} it is shown how the Derivative-free distributed nonlinear Kalman Filtering can be used for estimating the dynamics of distributed parameter systems as the ones described by 1D nonlinear PDEs of the wave type. In Section \ref{section 4 : Equivalence_between_Kalman_Filters_and_regressor_models} the equivalence between Kalman Filtering and regressor models is analyzed. In Section \ref{section 5 : Change_detection_using_the_local_statistical_approach} the local statistical approach is introduced as a systematic method for performing fault detection and isolation in dynamical systems. The method is proposed also for diagnosing faults in distributed parameter systems described by PDEs. In Section  \ref{section 6 : Simulation_tests} simulation tests are presented about the performance of the Derivative-free distributed nonlinear Kalman Filter in the problem of state estimation of the wave-type type partial differential estimation and about detecting and isolating parametric changes in such systems. Finally, in Section \ref{Section 7: Conclusions} concluding remarks are stated.

\section{Filtering using differential flatness theory and canonical forms} \label{section 2: Filtering_using_differential_flatness_theory_and_canonical_forms}

%%%%%%%%%%%%%%%%%%%%%%%%%%%%%%%%%

%@ \subsection{Definition of differentially flat systems}

%@ \noindent Differential flatness theory is currently a main direction in the analysis of nonlinear dynamical systems  %@ [\ref{Rud03}-\ref{Rig11}]. To deduce if a dynamical system is differentially flat, the following should be examined: %@ (i) the existence of the so-called flat output, i.e. a new variable which is expressed as a function of the system's %@ state variables. It should hold that the flat output and its derivatives should not be coupled in the form of an
%@ ordinary differential equation, (ii) the components of the system (i.e. state variables and control input) should be %@ expressed as functions of the flat output and its derivatives [\ref{Lev10}-\ref{FliMou99}].

\subsection{Conditions for applying the differential flatness theory} \label{subsection 3.3: diffeomorphism_canonical_form}

\noindent Next, a new filter will be developed, in accordance to the differential flatness theory. It will be shown that the filter can be efficiently used in the problem of state estimation in distributed parameter systems. First, the  generic class of nonlinear systems $\dot{x}=f(x,u)$ (including MIMO systems) is considered. Such systems can be transformed to the form of an affine in-the-input system by adding an integrator to each input [\ref{Lev10}],[\ref{BouBouZheBarKra11}]

\begin{equation} \label{affine_in_the_input_system}
\begin{tabular}{c}
$\dot{x}=f(x)+{\sum_{i=1}^m}g_i(x){u_i}$
\end{tabular}
\end{equation}

\noindent The following definitions are now used [\ref{Rig12a}]:\\

\noindent (i) Lie derivative:  ${L_f}h(x)$ stands for the Lie derivative ${L_f}h(x)=({\nabla}h)f$ and the repeated Lie derivatives are recursively defined as ${L_f^0}h=h \ \ \text{for} \ i=0$, ${L_f^i}h={L_f}{{L_f^{i-1}}h}=\nabla{L_f^{i-1}h}f \ \ \text{for} \ i=1,2,\cdots$. \\

\noindent (ii) Lie Bracket: ${ad_f^i}g$ stands for a Lie Bracket which is defined recursively as ${ad_f^i}g=[f,{ad_f^{i-1}}]g$ with ${ad_f^0}g=g$ and ${ad_f}g=[f,g]={{\nabla}g}f-{{\nabla}f}g$.  \\

\noindent If the system of Eq. (\ref{affine_in_the_input_system}) can be linearized by a diffeomorphism $z=\phi(x)$ and a static state feedback $u=\alpha(x)+\beta(x)v$ into the following form

\begin{equation} \label{system_transformed_in_canonical_form}
\begin{tabular}{c}
$\dot{z}_{i,j}=z_{i+1,j} \ \text{for} \ 1{\leq}j{\leq}m \ \text{and} \ 1{\leq}i{\leq}v_j-1$ \\
$\dot{z}_{v_{i,j}}={v_j}$
\end{tabular}
\end{equation}

\noindent with ${\sum_{j=1}^m}{v_j}=n$, then $y_j=z_{1,j}$ for $1{\leq}j{\leq}m$ are the 0-flat outputs which can be written as functions of only the elements of the state vector $x$. To define conditions for transforming the system of Eq. (\ref{affine_in_the_input_system}) into the canonical form described in Eq. (\ref{system_transformed_in_canonical_form}) the following theorem holds [\ref{BouBouZheBarKra11}] \\

\noindent \textit{Theorem}: For nonlinear systems described by Eq. (\ref{affine_in_the_input_system}) the following variables are defined: (i) $G_0=\text{span}[g_1,\cdots,g_m]$, (ii) $G_1=\text{span}[g_1,\cdots,g_m,ad_f{g_1},\cdots,ad_f{g_m}]$, $\cdots$
(k) $G_k=\text{span}\{{ad_f^j}{g_i} \ \text{for} \ 0{\leq}j{\leq}k, \ 1{\leq}i{\leq}m \}$.
Then, the linearization problem for the system of Eq. (\ref{affine_in_the_input_system})
can be solved if and only if: (1). The dimension of $G_i, \ i=1,\cdots,k$ is constant for $x{\in}X{\subseteq}R^n$ and for $1{\leq}i{\leq}n-1$, (2). The dimension of $G_{n-1}$ if of order $n$, (3). The distribution $G_k$ is involutive for each $1{\leq}k{\leq}{n-2}$.

\subsection{Transformation of MIMO systems into canonical forms} \label{subsection 3.4: transformation into the Brunovksy form}

\noindent It is assumed now that after defining the flat outputs of the initial MIMO nonlinear system and after expressing the system state variables and control inputs as functions of the flat output and of the associated derivatives, the system can be transformed in the Brunovsky canonical form:

\begin{equation} \label{MIMO_system_Brunovsky_canonical_form}
\begin{tabular}{l}
$\dot{x}_1=x_2$ \\
% $\dot{x}_2=x_3$ \\
$\cdots$ \\
$\dot{x}_{r_1-1}=x_{r_1}$ \\
$\dot{x}_{r_1}=f_1(x)+{\sum_{j=1}^p}{g_{1_j}(x)}u_j+d_1$\\
$$ \\
$\dot{x}_{r_1+1}=x_{r_1+2}$ \\
% $\dot{x}_{r_1+2}=x_{r_1+3}$ \\
$\cdots$ \\
$\dot{x}_{p-1}=x_{p}$ \\
$\dot{x}_{p}=f_p(x)+{\sum_{j=1}^p}{g_{p_j}(x)}u_j+d_p$\\
$$ \\
$y_1=x_1$ \\
% $y_2=x_2$ \\
$\cdots$ \\
$y_p=x_{n-r_p+1}$
\end{tabular}
\end{equation}

\noindent where $x=[x_1,\cdots,x_n]^T$ is the state vector of the transformed system (according to the differential flatness formulation), $u=[u_1,\cdots,u_p]^T$ is the set of control inputs, $y=[y_1,\cdots,y_p]^T$ is the output vector, $f_i$ are the drift functions and $g_{i,j}, \ i,j=1,2,\cdots,p$ are smooth functions corresponding to the control input gains, while $d_j$ is a variable associated to external disturbances. It holds that $r_1+r_2+\cdots+r_p=n$. Having written the initial nonlinear system into the canonical (Brunovsky) form it holds

\begin{equation} \label{compact_Brunovsky_form_dynamics}
\begin{tabular}{c}
$y_i^{(r_i)}=f_i(x)+{\sum_{j=1}^p}g_{ij}(x)u_j+d_j$
\end{tabular}
\end{equation}

\noindent Next the following vectors and matrices can be defined:
$f(x)=[f_1(x),\cdots,f_n(x)]^T$, $g(x)=[g_1(x),\cdots,g_n(x)]^T$, with  $g_i(x)=[g_{1i}(x),\cdots,g_{pi}(x)]^T$, $A=diag[A_1,\cdots,A_p], \ \  B=diag[B_1,\cdots,B_p]$, $C^T=diag[C_1,\cdots,C_p], \ \ d=[d_1,\cdots,d_p]^T$, where matrix $A$ has the MIMO canonical form, i.e. with block-diagonal elements

\begin{equation} \label{matrix_A_B_C_MIMO_canonical_form}
\begin{tabular}{c}
$A_i=\begin{pmatrix}
0 & 1 & \cdots & 0 \\
0 & 0 & \cdots & 0 \\
\vdots & \vdots & \cdots & \vdots \\
0 & 0 & \cdots & 1 \\
0 & 0 & \cdots & 0
\end{pmatrix}_{{r_i}\times{r_i}}$ \\
$ $\\
$B_i^T=\begin{pmatrix}
0 & 0 & \cdots 0 & 1
\end{pmatrix}_{1{\times}{r_i}}$ \\
$C_i=\begin{pmatrix}
1 & 0 & \cdots 0 & 0
\end{pmatrix}_{1{\times}{r_i}}$
\end{tabular}
\end{equation}

\noindent Thus, Eq. (\ref{compact_Brunovsky_form_dynamics}) can be written in state-space form

\begin{equation} \label{Brunovsky_form_MIMO_system}
\begin{tabular}{c}
$\dot{x}=Ax+Bv+B\tilde{d}$\\
$y={C}x$
\end{tabular}
\end{equation}

\noindent where the control input is written as $v=f(x)+g(x)u$. The system of Eq. (\ref{matrix_A_B_C_MIMO_canonical_form}) and Eq. (\ref{Brunovsky_form_MIMO_system}) is
in controller and observer canonical form.

\subsection{Derivative-free nonlinear Kalman Filtering} \label{subsection : flatness_based_control_2}

\noindent As mentioned above, for the system of Eq. (\ref{Brunovsky_form_MIMO_system}), state estimation is possible by applying the standard Kalman Filter. The system is first turned into discrete-time form using common discretization methods and then the recursion of the linear Kalman Filter described in Eq. (\ref{KF_meas_update}) and Eq. (\ref{KF_time_update}) is applied. \\

\noindent If the derivative-free Kalman Filter is used in place of the Extended Kalman Filter then in the EKF equations the following matrix substitutions should be performed: $J_{\phi}(k){\rightarrow}A_d$, $J_{\gamma}(k){\rightarrow}C_d$, where  matrices $A_d$ and $C_d$ are the discrete-time equivalents of matrices $A$ and $C$ which have been defined Eq. (\ref{Brunovsky_form_MIMO_system}) and which appear also in the measurement and time update of the standard Kalman Filter recursion. Matrices $A_d$ and $C_d$ can be computed using established discretization methods. Moreover, the covariance matrices $P(k)$ and $P^{-}(k)$ are the ones obtained from the linear Kalman Filter update equations given in Section \ref{Section 3: Estimation_of_nonlinear_wave_equations}.\\

\section{Estimation of nonlinear wave dynamics} \label{Section 3: Estimation_of_nonlinear_wave_equations}

\noindent The following nonlinear wave equation is considered

\begin{equation} \label{nonlinear_wave_equation}
\begin{tabular}{c}
${{{{\partial}^2}\phi} \over {{\partial}t^2}}=k{{{{\partial}^2}\phi} \over {{\partial}x^2}}+f(\phi)$
\end{tabular}
\end{equation}

\noindent Using the approximation for the partial derivative

\begin{equation} \label{nonlinear_wave_eq1}
\begin{tabular}{c}
${{{{\partial}^2}\phi} \over {{\partial}x^2}}{\simeq}={{\phi_{i+1}-2{\phi_i}+{\phi}_{i-1}} \over {{\Delta}x^2}}$
\end{tabular}
\end{equation}

\noindent and considering spatial measurements of variable $\phi$ along axis $x$ at points $x_0+i{\Delta}x, \ i=1,2,\cdots,N$ one has

\begin{equation}  \label{nonlinear_wave_eq2}
\begin{tabular}{c}
${{{{\partial}^2}\phi_i} \over {{\partial}t^2}}={K \over {{\Delta}x^2}}\phi_{i+1}-{{2K} \over {{\Delta}x^2}}\phi_i+{K \over {{\Delta}x}^2}\phi_{i-1}+f(\phi_i)$
\end{tabular}
\end{equation}

\noindent By considering the associated samples of $\phi$ given by $\phi_0, \phi_1, \cdots, \phi_N, \phi_{N+1}$
one has

\begin{equation}  \label{nonlinear_wave_eq3}
\begin{tabular}{c}
${{{{\partial}^2}\phi_1} \over {{\partial}t^2}}={K \over {{\Delta}x}^2}\phi_{2}-{{2K} \over {{\Delta}x}^2}\phi_1+{K \over {{\Delta}x}^2}\phi_{0}+f(\phi_1)$ \\
${{{{\partial}^2}\phi_2} \over {{\partial}t^2}}={K \over {{\Delta}x}^2}\phi_{3}-{{2K} \over {{\Delta}x}^2}\phi_2+{K \over {{\Delta}x}^2}\phi_{1}+f(\phi_2)$ \\
${{{{\partial}^2}\phi_3} \over {{\partial}t^2}}={K \over {{\Delta}x}^2}\phi_{4}-{{2K} \over {{\Delta}x}^2}\phi_3+{K \over {{\Delta}x}^2}\phi_{2}+f(\phi_3)$ \\
$\cdots$ \\
${{{{\partial}^2}\phi_{N-1}} \over {{\partial}t^2}}={K \over {{\Delta}x}^2}\phi_{N}-{{2K} \over {{\Delta}x}^2}\phi_{N-1}+{K \over {{\Delta}x}^2}\phi_{N-2}+f(\phi_{N-1})$ \\
${{{{\partial}^2}\phi_{N}} \over {{\partial}t^2}}={K \over {{\Delta}x}^2}\phi_{N+1}-{{2K} \over {{\Delta}x}^2}\phi_{N}+{K \over {{\Delta}x}^2}\phi_{N-1}+f(\phi_{N})$
\end{tabular}
\end{equation}

\noindent By defining the following state vector

\begin{equation} \label{state_vector_eq}
\begin{tabular}{c}
$x^T=\begin{pmatrix}
\phi_1,\phi_2,\cdots,\phi_N
\end{pmatrix}$
\end{tabular}
\end{equation}

\noindent one obtains the following state-space description

\begin{equation} \label{state_space_description1}
\begin{tabular}{c}
$\ddot{x}_1={K \over {{\Delta}x}^2}x_2-{{2K} \over {{\Delta}x}^2}{x_1}+{K \over {{\Delta}x}^2}{\phi_0}+f(x_1)$ \\
$\ddot{x}_2={K \over {{\Delta}x}^2}x_3-{{2K} \over {{\Delta}x}^2}{x_2}+{K \over {{\Delta}x}^2}{x_1}+f(x_2)$ \\
$\ddot{x}_3={K \over {{\Delta}x}^2}x_4-{{2K} \over {{\Delta}x}^2}{x_3}+{K \over {{\Delta}x}^2}{x_2}+f(x_3)$ \\
$\cdots$ \\
$\ddot{x}_{N-1}={K \over {{\Delta}x}^2}x_{N}-{{2K} \over {{\Delta}x}^2}{x_{N-1}}+{K \over {{\Delta}x}^2}{x_{N-2}}+f(x_{N-1})$\\
$\ddot{x}_{N}={K \over {{\Delta}x}^2}\phi_{N+1}-{{2K} \over {{\Delta}x}^2}{x_{N}}+{K \over
{{\Delta}x}^2}{x_{N-1}}+f(x_{N})$
\end{tabular}
\end{equation}

\noindent Next, the following state variables are defined

\begin{equation} \label{new_state_variables}
\begin{tabular}{c}
$y_{1,i}=x_i$ \\
$y_{2,i}=\dot{x}_i$
\end{tabular}
\end{equation}

\noindent and the state-space description of the system becomes as follows

\begin{equation} \label{state_space_description_v2}
\begin{tabular}{c}
$\dot{y}_{1,1}=y_{2,1}$ \\
$\dot{y}_{2,1}={K \over {{\Delta}x}^2}y_{1,2}-{{2K} \over {{\Delta}x}^2}y_{1,1}+{K \over {{\Delta}x}^2}\phi_0+f(y_{1,1})$ \\
$\dot{y}_{1,2}=y_{2,2}$ \\
$\dot{y}_{2,2}={K \over {{\Delta}x}^2}y_{1,3}-{{2K} \over {{\Delta}x}^2}y_{1,2}+{K \over {{\Delta}x}^2}y_{1,1}+f(y_{1,2})$ \\
$\dot{y}_{1,3}=y_{2,3}$ \\
$\dot{y}_{2,3}={K \over {{\Delta}x}^2}y_{1,4}-{{2K} \over {{\Delta}x}^2}y_{1,3}+{K \over {{\Delta}x}^2}y_{1,2}+f(y_{1,3})$ \\
$\cdots$ \\
$\cdots$ \\
$\dot{y}_{1,N-1}=y_{2,N-1}$ \\
$\dot{y}_{2,N-1}={K \over {{\Delta}x}^2}y_{1,N}-{{2K} \over {{\Delta}x}^2}y_{1,N-1}+{K \over {{\Delta}x}^2}y_{1,N-2}+f(y_{1,N-1})$ \\
$\dot{y}_{1,N}=y_{2,N}$ \\
$\dot{y}_{2,N}={K \over {{\Delta}x}^2}\phi_{N+1}-{{2K} \over {{\Delta}x^2}}y_{1,N}+{K \over {\Delta}x}y_{1,N-1}+f(y_{1,N})$
\end{tabular}
\end{equation}

\noindent The dynamical system described in Eq. (\ref{state_space_description_v2}) is a differentially flat one with flat output defined as the vector $\tilde{y}=[y_{1,1},y_{1,2},\cdots,y_{1,N}]$. Indeed all state variables can be written as functions of the flat output and its derivatives. \\

\noindent Moreover, by defining the new control inputs

\begin{equation} \label{modified_controlled_inputs}
\begin{tabular}{c}
$v_1={K \over {{\Delta}x}^2}y_{1,2}-{{2K} \over {{\Delta}x^2}}y_{1,1}+{K \over {{\Delta}x^2}}\phi_0+f(y_{1,1})$\\
$v_2={K \over {{\Delta}x}^2}y_{1,3}-{{2K} \over {{\Delta}x^2}}y_{1,2}+{K \over {{\Delta}x^2}}y_{1,1}+f(y_{1,2})$\\
$v_3={K \over {{\Delta}x}^2}y_{1,4}-{{2K} \over {{\Delta}x^2}}y_{1,3}+{K \over {{\Delta}x^2}}y_{1,2}+f(y_{1,3})$\\
$\cdots$ \\
$v_{N-1}={K \over {{\Delta}x}^2}y_{1,N}-{{2K} \over {{\Delta}x}^2}y_{1,N-1}+{{K \over {\Delta}x}^2}y_{1,N-2}+f(y_{1,N-1})$\\
$v_N={K \over {{\Delta}x}^2}\phi_{N+1}-{{2K} \over {{\Delta}x}^2}y_{1,N}+{K \over {{\Delta}x}^2}y_{1,N-1}+f(y_{1,N})$
\end{tabular}
\end{equation}

\noindent the following state-space description is obtained

\begin{equation}  \label{state_space_description_canonical_form}
\begin{tabular}{c}
$\begin{pmatrix}
\dot{y}_{1,1} \\
\dot{y}_{2,1} \\
\dot{y}_{1,2} \\
\dot{y}_{2,2} \\
\cdots \\
\dot{y}_{1,N-1} \\
\dot{y}_{2,N-1} \\
\dot{y}_{1,N} \\
\dot{y}_{2,N}
\end{pmatrix}=
\begin{pmatrix}
0 \ 1 \ 0 \ 0 \ \cdots \ 0 \ 0 \ 0 \ 0 \\
0 \ 0 \ 0 \ 0 \ \cdots \ 0 \ 0 \ 0 \ 0 \\
0 \ 0 \ 0 \ 1 \ \cdots \ 0 \ 0 \ 0 \ 0 \\
0 \ 0 \ 0 \ 0 \ \cdots \ 0 \ 0 \ 0 \ 0 \\
0 \ 0 \ 0 \ 0 \ \cdots \ 0 \ 0 \ 0 \ 0 \\
0 \ 0 \ 0 \ 0 \ \cdots \ 0 \ 0 \ 0 \ 0 \\
\cdots \ \cdots \ \cdots \ \cdots \  \cdots \ \cdots  \\
0 \ 0 \ 0 \ 0 \ \cdots \ 0 \ 1 \ 0 \ 0 \\
0 \ 0 \ 0 \ 0 \ \cdots \ 0 \ 0 \ 0 \ 0 \\
0 \ 0 \ 0 \ 0 \ \cdots \ 0 \ 0 \ 0 \ 1 \\
0 \ 0 \ 0 \ 0 \ \cdots \ 0 \ 0 \ 0 \ 0
\end{pmatrix}
\begin{pmatrix}
y_{1,1} \\
y_{2,1} \\
y_{1,2} \\
y_{2,2} \\
\cdots \\
y_{1,N-1} \\
y_{2,N-1} \\
y_{1,N} \\
y_{2,N}
\end{pmatrix}+$\\
$$\\
$+\begin{pmatrix}
0 \ 0 \ 0 \ \cdots \ 0 \ 0 \\
1 \ 0 \ 0 \ \cdots \ 0 \ 0 \\
0 \ 0 \ 0 \ \cdots \ 0 \ 0 \\
0 \ 1 \ 0 \ \cdots \ 0 \ 0 \\
0 \ 0 \ 0 \ \cdots \ 0 \ 0 \\
0 \ 0 \ 1 \ \cdots \ 0 \ 0 \\
\cdots \ \cdots \ \cdots \ \cdots \\
0 \ 0 \ 0 \ \cdots \ 0 \ 0 \\
0 \ 0 \ 0 \ \cdots \ 1 \ 0 \\
0 \ 0 \ 0 \ \cdots \ 0 \ 0 \\
0 \ 0 \ 0 \ \cdots \ 0 \ 1
\end{pmatrix}
\begin{pmatrix}
v_1 \\
v_2 \\
v_3 \\
\cdots \\
v_{N-1} \\
v_N
\end{pmatrix}$
\end{tabular}
\end{equation}

\noindent By selecting measurements from a subset of points $x_j \ j{\in}[1,2,\cdots,m]$, the associated observation (measurement) equation becomes

\begin{equation} \label{measurement_equation}
\begin{tabular}{c}
$\begin{pmatrix}
z_1 \\
z_2 \\
\cdots \\
z_m
\end{pmatrix}=
\begin{pmatrix}
1 \ 0 \ 0 \ \cdots \ 0 \ 0 \\
0 \ 0 \ 0 \ \cdots \ 0 \ 0 \\
\cdots \ \cdots \ \cdots \ \cdots \\
0 \ 0 \ 0 \ \cdots \ 1 \ 0 \\
0 \ 0 \ 0 \ \cdots \ 0 \ 0
\end{pmatrix}
\begin{pmatrix}
y_{1,1} \\
y_{2,1} \\
y_{1,2} \\
y_{2,2} \\
\cdots \\
y_{1,N} \\
y_{2,N}
\end{pmatrix}$
\end{tabular}
\end{equation}

\noindent Thus, in matrix form one has the following state-space description of the system

\begin{equation} \label{state_space_description_matrix_form}
\begin{tabular}{c}
$\dot{\tilde{y}}=A\tilde{y}+Bv$ \\
$\tilde{z}=C\tilde{y}$
\end{tabular}
\end{equation}

\noindent Denoting  $a={K \over {Dx^2}}$ and $b=-{{2K} \over {Dx^2}}$, the initial description of the system given in Eq. (\ref{state_space_description_canonical_form}) is rewritten as follows

\begin{equation}  \label{state_space_description_canonical_form_v2}
\begin{tabular}{c}
$\begin{pmatrix}
\dot{y}_{1,1} \\
\dot{y}_{2,1} \\
\dot{y}_{1,2} \\
\dot{y}_{2,2} \\
\cdots \\
\dot{y}_{1,N-1} \\
\dot{y}_{2,N-1} \\
\dot{y}_{1,N} \\
\dot{y}_{2,N}
\end{pmatrix}=
\begin{pmatrix}
0 \ 1 \ 0 \ 0 \ 0 \ 0 \ 0 \ \cdots \ 0 \ 0 \ 0 \ 0 \ 0 \ 0 \\
b \ 0 \ a \ 0 \ 0 \ 0 \ 0 \ \cdots \ 0 \ 0 \ 0 \ 0 \ 0 \ 0 \\
0 \ 0 \ 0 \ 1 \ 0 \ 0 \ 0 \ \cdots \ 0 \ 0 \ 0 \ 0 \ 0 \ 0 \\
a \ 0 \ b \ 0 \ a \ 0 \ 0 \ \cdots \ 0 \ 0 \ 0 \ 0 \ 0 \ 0 \\
0 \ 0 \ 0 \ 0 \ 0 \ 1 \ 0 \ \cdots \ 0 \ 0 \ 0 \ 0 \ 0 \ 0 \\
0 \ 0 \ a \ 0 \ b \ 0 \ a \ \cdots \ 0 \ 0 \ 0 \ 0 \ 0 \ 0 \\
\cdots \ \cdots \ \cdots \ \cdots \  \cdots \ \cdots \ \cdots   \\
0 \ 0 \ 0 \ 0 \ 0 \ 0 \ 0 \ \cdots \ 0 \ 0 \ 0 \ 1 \ 0 \ 0 \\
0 \ 0 \ 0 \ 0 \ 0 \ 0 \ 0 \ \cdots \ a \ 0 \ b \ 0 \ a \ 0 \\
0 \ 0 \ 0 \ 0 \ 0 \ 0 \ 0 \ \cdots \ 0 \ 0 \ 0 \ 0 \ 0 \ 1 \\
0 \ 0 \ 0 \ 0 \ 0 \ 0 \ 0 \ \cdots \ 0 \ 0 \ a \ 0 \ b \ 0
\end{pmatrix}
\begin{pmatrix}
y_{1,1} \\
y_{2,1} \\
y_{1,2} \\
y_{2,2} \\
\cdots \\
y_{1,N-1} \\
y_{2,N-1} \\
y_{1,N} \\
y_{2,N}
\end{pmatrix}+$\\
$$ \\
$+\begin{pmatrix}
0 \ 0 \ 0 \ \cdots \ 0 \ 0 \\
1 \ 0 \ 0 \ \cdots \ 0 \ 0 \\
0 \ 0 \ 0 \ \cdots \ 0 \ 0 \\
0 \ 1 \ 0 \ \cdots \ 0 \ 0 \\
0 \ 0 \ 0 \ \cdots \ 0 \ 0 \\
0 \ 0 \ 1 \ \cdots \ 0 \ 0 \\
\cdots \ \cdots \ \cdots \ \cdots \\
0 \ 0 \ 0 \ \cdots \ 0 \ 0 \\
0 \ 0 \ 0 \ \cdots \ 1 \ 0 \\
0 \ 0 \ 0 \ \cdots \ 0 \ 0 \\
0 \ 0 \ 0 \ \cdots \ 0 \ 1
\end{pmatrix}
\begin{pmatrix}
v_1 \\
v_2 \\
v_3 \\
\cdots \\
v_{N-1} \\
v_N
\end{pmatrix}$
\end{tabular}
\end{equation}

\noindent The associated control inputs are defined as

\begin{equation}    \label{control_inputs_2}
\begin{tabular}{c}
$v_1={K \over {{\Delta}x^2}}\phi_0+f(y_{1,1})$ \\
$v_2=f(y_{1,2})$\\
$v_3=f(y_{1,3})$\\
$\cdots$ \\
$v_{N-1}=f(y_{1,N-1})$ \\
$v_{N}=f(y_{1,N})$
\end{tabular}
\end{equation}

\noindent By selecting measurements from a subset of points $x_j \ j{\in}[1,2,\cdots,m]$, the associated observation (measurement) equation remains as in Eq. (\ref{measurement_equation}), i.e.

\begin{equation} \label{measurement_equation_2}
\begin{tabular}{c}
$\begin{pmatrix}
z_1 \\
z_2 \\
\cdots \\
z_m
\end{pmatrix}=
\begin{pmatrix}
1 \ 0 \ 0 \ \cdots \ 0 \ 0 \\
0 \ 0 \ 0 \ \cdots \ 0 \ 0 \\
\cdots \ \cdots \ \cdots \ \cdots  \\
0 \ 0 \ 0 \ \cdots \ 1 \ 0 \\
0 \ 0 \ 0 \ \cdots \ 0 \ 0
\end{pmatrix}
\begin{pmatrix}
y_{1,1} \\
y_{2,1} \\
y_{1,2} \\
y_{2,2} \\
\cdots \\
y_{1,N} \\
y_{2,N}
\end{pmatrix}$
\end{tabular}
\end{equation}

\noindent For the linear description of the system in the form of Eq. (\ref{state_space_description_matrix_form}) one can perform estimation using the standard Kalman Filter recursion. The discrete-time Kalman filter can be decomposed into two parts: i) time update (prediction stage), and ii) measurement update (correction stage).\\

\noindent \textit{measurement update}:

\begin{equation} \label{KF_meas_update}
\begin{tabular}{l}
$K(k)={P^{-}(k)}{C^T}[C{\cdot}P^{-}(k){C^T}+R]^{-1}$ \\
${\hat{y}(k)}={\hat{y}^{-}(k)}+K(k)[z(k)-C{\hat{y}^{-}(k)}]$ \\
$P(k)=P^{-}(k)-K(k)CP^{-}(k)$
\end{tabular}
\end{equation}

\noindent \textit{time update}:

\begin{equation} \label{KF_time_update}
\begin{tabular}{c}
$P^{-}(k+1)={A(k)}P(k)A^T(k)+Q(k)$\\
${\hat{y}^{-}(k+1)}=A(k)\hat{y}(k)+B(k)u(k)$
\end{tabular}
\end{equation}

\noindent Therefore, by taking measurements of $\phi(x,t)$ at time instant $t$ at a small number of measuring points $j=1,\cdots,n_1$ it is possible to estimate the complete state vector, i.e. to get values of $\phi$ in a mesh of points that covers efficiently the variations of $\phi(x,t)$. By processing a sequence of output measurements of the system, one can obtain local estimates of the state vector $\hat{y}$. The measuring points (active sensors) can vary in time provided that the observability criterion for the state-space model of the PDE holds.\\

\noindent \textit{Remark}: The proposed derivative-free nonlinear Kalman Filter is of improved precision because unlike other nonlinear filtering schemes, e.g. the Extended Kalman Filter it does not introduce cumulative numerical errors due to approximative linearization of the system's dynamics. Besides it is computationally more efficient (faster) because it does not require to calculate Jacobian matrices and partial derivatives.

%@+ \begin{figure}[htb]
%@+ \begin{center}
%@+ \rotatebox{-90}{\epsfig{file=Grid.eps, width=60mm,
%@+ height=70mm}}
%@+ \end{center}
%@+ \caption{Grid points for measuring $\phi(x,t)$} \label{figure : Grid_points}
%@+ \end{figure}

\begin{figure}[htb]
\begin{center}
\rotatebox{-90}{\epsfig{file=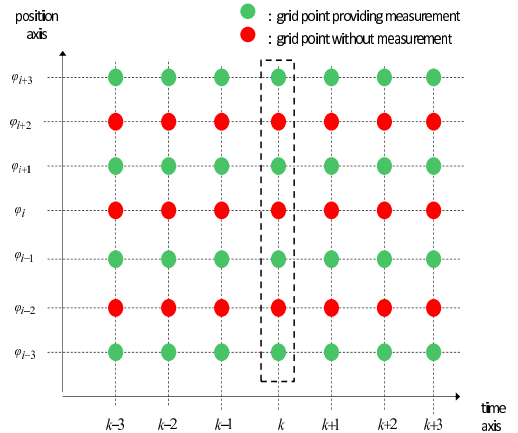, width=60mm,
height=70mm}}
\end{center}
\caption{Grid points for measuring $\phi(x,t)$} \label{figure : Grid_points}
\end{figure}

\section{Equivalence between Kalman filters and regressor models}  \label{section 4 : Equivalence_between_Kalman_Filters_and_regressor_models}

\subsection{Equivalence between the standard Kalman Filter and linear regressor models}

\noindent For fault diagnosis purposes it is convenient to turn the Kalman Filter model of distributed parameter systems into equivalent ARMAX (autoregressive moving average model with auxiliary input) models. An ARMAX model is an input-output model of the form

\begin{equation} \label{ARMAX_model}
\begin{tabular}{c}
$A(z){Y_k}=C(z)U_k+B(z)\{\epsilon_k\}$
\end{tabular}
\end{equation}

\noindent $A$, $B$, $C$ are polynomial matrices in the backwards shift operator $z^{-1}$:

\begin{equation} \label{ARMA_model_components}
\begin{tabular}{c}
$A(z)=A_0-{\sum_{i=1}^p}{A_i}{z^{-i}}$ \\
$B(z)={\sum_{j=0}^q}{B_j}{z^{-j}}$ \\
$C(z)={\sum_{l=1}^l}{C_l}z^{-l}$
\end{tabular}
\end{equation}

\noindent such that $A$ has non-singular constant term $A_0$ and where ${\epsilon_k}$ is a white noise sequence with covariance matrix $R$. A state-space model and particularly the Kalman Filter estimator can be written in the form of an ARMAX model. For linear systems, the Kalman Filter (for the single-input case) can be written in the form [\ref{BasNik93}]

\begin{equation} \label{Kalman_Filter1}
\begin{tabular}{c}
$\begin{pmatrix}
\hat{x}_1(k+1) \\
\hat{x}_2(k+1)\\
\cdots \\
\cdots \\
\hat{x}_n(k+1)
\end{pmatrix}=
\begin{pmatrix}
\alpha_1 & 1 & 0 & \cdot & 0 \\
\cdot & \cdot & \cdot & \cdot & \cdot \\
\cdot & \cdot & \cdot & \cdot & \cdot \\
0 & \cdot & \cdot & \cdot & 1 \\
\alpha_n & \cdot & \cdot & \cdot & 0
\end{pmatrix}
\begin{pmatrix}
\hat{x}_1(k) \\
\hat{x}_2(k)\\
\cdots \\
\cdots \\
\hat{x}_n(k)
\end{pmatrix}+$\\
$$\\
$+\begin{pmatrix}
g_1 \\
\cdot \\
\cdot \\
\cdot \\
g_n
\end{pmatrix}{U_k}+
\begin{pmatrix}
\kappa_1(k) \\
\cdot \\
\cdot \\
\cdot \\
\kappa_n(k)
\end{pmatrix}
\epsilon_k
$
\end{tabular}
\end{equation}

\begin{equation} \label{Kalman_Filter2}
\begin{tabular}{c}
$\hat{Y}(k)=
\begin{pmatrix}
1 & 0 & \cdot & \cdot & 0
\end{pmatrix}
\hat{X}(k-1)+J{U_k}+{\epsilon_k}$
\end{tabular}
\end{equation}

\noindent Using successive substitutions this can be rewritten as a time-varying ARMAX model:

\begin{equation} \label{ARMAX_model_Kalman_Filter_1}
A(z){Y_k}=C(z){U_k}+B(k,z){\epsilon_k}
\end{equation}

\noindent where

\begin{equation} \label{ARMAX_model_Kalman_Filter_2}
\begin{tabular}{c}
$A(z)=1-{\alpha_1}{z^{-1}}-\cdots-{\alpha_n}{z^{-n}}$ \\
$C(z)={g_1}{z^{-1}}+\cdots+{g_n}{z^{-n}}+JA(z^{-1})$ \\
$B(\kappa,z)=1+[\kappa_1(k-1)-\alpha_1]{z^{-1}}+\cdots+$\\
$+[\kappa_n(k-n)-\alpha_n]{z^{-n}}$
\end{tabular}
\end{equation}

\noindent Matrix $B(\kappa,z)$ is time-varying because the Kalman Filter gain $K_k$ is time-varying. But, under the conditions of the stability theorem, $K$ and $B$ are asymptotically constant. Thus, the ARMAX description of the Kalman Filter becomes

\begin{equation} \label{ARMAX_model_Kalman_Filter_3}
\begin{tabular}{c}
$A(z){Y_k}=C(z)U_k+B(z)\epsilon_k$
\end{tabular}
\end{equation}

\noindent This approach also holds for multi-input systems and one can transform again the state-space representation into an ARMAX model. \\

\section{Change detection with the local statistical approach}  \label{section 5 : Change_detection_using_the_local_statistical_approach}

\subsection{The global $\chi^2$ test for change detection}

\begin{figure}[htb]
\begin{center}
\rotatebox{-90}{\epsfig{file=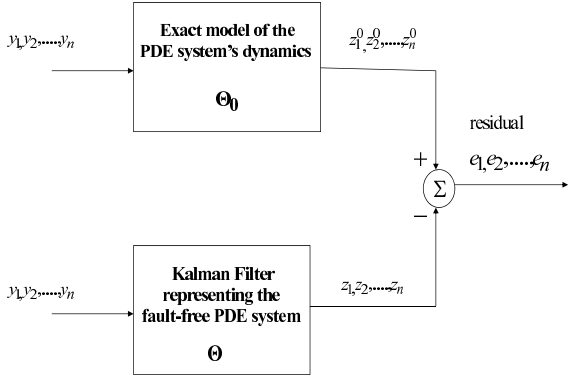, width=70mm,
height=90mm}} \caption{Residual between the output of the distributed parameter system and the Kalman Filter that models the PDE dynamics in the fault-free case} \label{fig : residual}
\end{center}
\end{figure}

\noindent Fault diagnosis for distributed parameter systems is based on the processing of the residuals, i.e. of the differences between the outputs of the PDE model and the outputs of the associated Kalman Filter. First, the residual $e_i$ is defined as the difference between the Kalman filter output $\hat{y}_i$ and the physical system output $y_i$, i.e. $e_i=\hat{y}_i-y_i$. It is also acceptable to define the residual as the difference between the Kalman Filter  output and the exact model output, where the exact model replaces the physical system and has the same number of parameters as the Kalman Filter (see Fig. \ref{fig : residual}). The partial derivative of the residual square is:

\begin{equation} \label{partial_derivative_residual}
\begin{tabular}{c}
$H(\theta,y_i)={1 \over 2}{{{\partial}{e_i^2}} \over {{\partial}{\theta}}}={e_i}{{{\partial}{\hat{y}_i}} \over {{\partial}{\theta}}}$
\end{tabular}
\end{equation}

\noindent The local statistical approach to fault diagnosis is a statistical method of fault diagnosis which can be used for consistency checking of the fuzzy Kalman Filter. Based on a small parametric disturbance assumption, the proposed FDI method aims at transforming complex detection problems concerning a parameterized stochastic process into the problem of monitoring the mean of a Gaussian vector. The local statistical approach consists of two stages : i) the global   test which indicates the existence of a change in some parameters of the distributed parameter system, ii) the diagnostics tests (sensitivity or min-max) which isolate the parameter affected by the change. The method's stages are analyzed first, following closely the method presented in [\ref{BasNik93}],[\ref{ZhaBasBen98}]. \\

%@+ the proposed method is based on the definition of the residual $e_i$
%@+ described as the difference between the output from the ARMAX (Kalman Filter) model associated with the changed
%@+ dynamics of the PDE and the output of the  ARMAX (Kalman Filter) model associated with the unchanged dynamics of the %@+ PDE.

\noindent As shown in Fig. \ref{fig : residual} the concept of this FDI technique is as follows: there is a Kalman Filter (ARMAX) model that represents the unchanged PDE dynamics. At each time instant the output of the aforementioned reference model is compared to the output of the Kalman Filter (ARMAX) model that represents the changed dynamics of the PDE. The difference $e_i$ between these two output measurements is called residual. The statistical processing of a sufficiently large number of residuals through an FDI method provides an index-variable that is compared against a fault threshold and which can give early indication about deviation of the PDE model from its fault-free condition. Under certain conditions (detectability of changes) the proposed FDI method enables also fault isolation, i.e. to identify the source of fault within the distributed parameter system. \\

\noindent Considering the representation of the Kalman Filter as an ARMAX model, the partial derivative of the residual square is:

\begin{equation} \label{partial_derivative_residual}
\begin{tabular}{c}
$H(\theta,\hat{y}_i)={1 \over 2}{{{\partial}{e_i^2}} \over {{\partial}{\theta}}}={e_i}{{{\partial}{\hat{y}_i}} \over {{\partial}{\theta}}}$
\end{tabular}
\end{equation}

\noindent where $\theta$ is the vector of model's parameters. The vector $H$ having as elements the above $H(\theta,\hat{y}_i)$ is called primary residual. The gradient of the output with respect to the ARMAX model parameters are given by

\begin{equation} \label{derivative_output_weight}
{{{\partial}{\hat{y}}} \over {{\partial}w_i}}={{x_i}}
\end{equation}

%@+ \noindent The gradient with respect to the center $c_i^l$ has been given in Eq. (\ref{gradient_centers}) while the %@+ gradient with respect to the spread $v_i^l$ has been given in Eq. (\ref{gradient_spreads}).

\noindent Next, having calculated the partial derivatives of Eq.(\ref{derivative_output_weight}), the rows of the Jacobian matrix $J$ are found by

\begin{equation} \label{Jacobian}
{J(\theta_0,\hat{y}_k)}={\partial {\hat{y}_k(\theta)} \over {\partial{\theta}} } {\Bigg\vert}_{\theta={\theta_0}}
\end{equation}

\noindent where $\theta_0$ represents the nominal value of the parameters. The problem of change detection with the $\chi^2$ test consists of monitoring a change in the mean of the Gaussian variable which for the one-dimensional parameter vector $\theta$ is formulated as

\begin{equation} \label{Gaussian_variable}
X={1 \over {\sqrt{N}}}{\sum_{i=1}^N}{e_k}{{{\partial}{\hat{y}_k}} \over {{\partial}{\theta}}}{\sim}\emph{N}(\mu,\sigma^2)
\end{equation}

\noindent where $\hat{y}_k$ is the output of the Kalman Filter (ARMAX) model generated by the input pattern $x_k$, $e_k$ is the associated residual and $\theta$ is the vector of the model's parameters. It is noted that $X$ is the monitored parameter for the FDI test, which means that when the mean value of $X$ is $0$ the system is in the fault-free condition, while when the mean value of $X$ has moved away from $0$ the system is in a faulty condition. For a  multivariable parameter vector $\theta$ should hold $X{\sim}{\emph{N}}(M{{\delta}\theta}, S)$, where $S$ denotes the covariance matrix of $X$. In order to decide if the system (Kalman Filter) is in fault-free operating conditions, given a set of data of $N$ measurements, let $\theta_{*}$ be the value of the parameters vector $\mu$ minimizing the RMSE. The notation is introduced only for the convenience of problem formulation, and its actual value does not need to be known. Then the model validation problem amounts to make a decision between the two hypotheses:

\begin{equation}\label{hypothesis_test}
\begin{tabular}{c}
$H_0: \ \theta_*=\theta_0$ \\
$H_1: \ \theta_*=\theta_0+{1 \over \sqrt{N}}{\delta}\theta$
\end{tabular}
\end{equation}

\noindent where ${\delta}{\theta}{\neq}0$. It is known from the Central Limit Theorem that for a large data sample, the normalized residual given by Eq.(\ref{Gaussian_variable}) asymptotically follows a Gaussian distribution when $\text{N}{\rightarrow}\infty$ [\ref{ZhaBasBen98}-\ref{BasBenZha96}]. More specifically, the hypothesis that has to be tested is:

\begin{eqnarray*}
& H_0 : X \sim \emph{N}(0,S)  \\
& H_1 : X \sim \emph{N}(M{\delta}{\theta},S)
\end{eqnarray*}

\noindent where $M$ is the sensitivity matrix (see Eq. (\ref{sensitivity_matrix})),
${\delta}{\theta}$ is the change in the parameters' vector and $S$ is the convariance matrix (see Eq.
(\ref{covariance_matrix})). The product $M{{\delta}{\theta}}$  denotes the new center of the
monitored Gaussian variable $X$, after a change on the system's parameter $\theta$.
The sensitivity matrix $M$ of ${1 \over {\sqrt{N}}}X$ is defined as the mean value
of the partial derivative with respect to $\theta$ of the primary residual defined in
Eq. \ref{partial_derivative_residual}, i.e. $E\{{{\partial} \over {{\partial}\theta}}H(\theta,\hat{y}_k)\}$
and is approximated by [8]:\\

\begin{equation} \label{sensitivity_matrix}
\begin{tabular}{c}
$M(\theta_0){\simeq}{{\partial} \over {{\partial}{\theta}}}{1 \over N}{\sum_{k=1}^N}H(\theta_0,\hat{y}_k){\simeq}{1 \over N}{J^T}J$
\end{tabular}
\end{equation}

\noindent The covariance matrix $S$ is defined as $E\{H(\theta,y_k){H^T}(\theta,\hat{y}_{k+m})\}$, $m=0,\pm1,\cdots$ and is approximated by [\ref{BasBenZha96}]:

\begin{equation} \label{covariance_matrix}
\begin{tabular}{l}
$S={\simeq}{\sum_{k=1}^N}[{H(\theta_0,\hat{y}_k)}{H^T(\theta_0,\hat{y}_k)}]+$\\
$+{\sum_{m=1}^I}{1 \over {N-m}}{\sum_{k=1}^{N-m}}[H(\theta_0,\hat{y}_k)H^T(\theta_0,\hat{y}_{k+m})+$\\
$+H(\theta_0,\hat{y}_{k+m})H^T(\theta_0,\hat{y}_k)]$
\end{tabular}
\end{equation}

\noindent \\

\noindent where an acceptable value for $I$ is 3. The decision tool is the likelihood ratio $s(X)=ln{{p_{\theta_1(x)}} \over {p_{\theta_0(x)}}}$, where $p_{\theta_1}(X)=e^{[X-\mu(X)]^T{S^{-1}}[X-\mu(X)]}$ and $p_{\theta_0}(X)=e^{{X^T}{S^{-1}}X}$. The center of the Gaussian distribution of the changed system is denoted as $\mu(X)=M{\delta}{\theta}$ where $\delta{\theta}$ is the change in the parameters vector. The \textit{Generalized Likelihood Ratio} (GLR) is calculated by maximizing the likelihood ratio with respect to
${\delta}{\theta}$ [\ref{BasBenZha96}]. This means that the most likely case of parameter change is taken into account. This gives the global $\chi^2$ test $t$:

\begin{equation} \label{chi2_test}
\begin{tabular}{c}
$t={X^T}{S^{-1}}M({M^T}{S^{-1}}M)^{-1}{M^T}S^{-1}X$
\end{tabular}
\end{equation}

\noindent Since $X$ asymptotically follows a Gaussian distribution, the statistics defined in Eq. (\ref{chi2_test}) follows a $\chi^2$ distribution with $n$ degrees of freedom. Mapping the change detection problem to this $\chi^2$ distribution enables the choice of the change threshold. Assume that the desired probability of false alarm is $\alpha$ then the change threshold $\lambda$ should be chosen from the relation

\begin{equation} \label{alarm_threshold}
\begin{tabular}{c}
${\int_{\lambda}^{\infty}}{\chi_n^2}(s)ds=\alpha$,
\end{tabular}
\end{equation}

\noindent where ${\chi_n^2}(s)$ is the probability density function (p.d.f.) of a variable that follows the $\chi^2$ distribution with $n$ degrees of freedom.

\subsection{Statistical fault isolation with the sensitivity test}

\noindent Fault isolation is needed to identify the source of faults and parametric changes in the PDE system. This means that the fault diagnosis method should be able to find out, among the complete set of parameters of the PDE, which are the ones that are subject to change with respect to their nominal values. A first approach to change isolation is to focus only on a subset of the parameters while considering that the rest of the parameters unchanged [\ref{BasBenZha96}]. The parameters vector $\eta$ can be written as $\eta=[\phi,\psi]^T$, where $\phi$ contains those parameters to be subject to the isolation test ,while $\psi$ contains those parameters to be excluded from the isolation test. $M_{\phi}$ contains the columns of the sensitivity matrix $M$ which are associated with the parameters subject to the isolation test. Similarly $M_{\psi}$ contains the columns of $M$ that are associated with the parameters to be excluded from the sensitivity test.\\

\noindent Assume that among the parameters $\eta$, it is only the subset $\phi$ that is suspected to have undergone a change. Thus $\eta$ is restricted to $\eta=[\phi,0]^T$. The associated columns of the sensitivity matrix are given by $M_{\phi}$ and the mean of the Gaussian to be monitored is $\mu=M_{\phi}{\phi}$, i.e.

\begin{equation}
\mu=MA{\phi},  \ \
A=[0, I]^T
\end{equation}

\noindent Matrix $A$ is used to select the parameters that will be subject to the fault isolation test. The rows of
$A$ correspond to the total set of parameters while the columns of A correspond only to the parameters selected for
the test. Thus the fault diagnosis ($\chi^2$) test of Eq. (\ref{chi2_test}) can be restated as:

\begin{equation} \label{chi2_test_sensitivity}
\begin{tabular}{c}
$t_{\phi}={X^T}S^{-1}{M_{\phi}}({M_{\phi}^T}{S^{-1}}M_{\phi})^{-1}{M_{\phi}^T}{S^{-1}X}$
\end{tabular}
\end{equation}

\subsection{Statistical fault isolation with the min-max test\\}

\noindent In this approach the aim is to find a statistic that will be able to detect a change on the part $\phi$ of the
parameters vector $\eta$ and which will be robust to a change in the non observed part $\psi$ [\ref{BasBenZha96}]. Assume the
vector partition $\eta=[\phi,\psi]^T$ . The following notation is used:

\begin{equation}
{M^T}{S^{-1}}M=
\begin{pmatrix}
I_{\varphi\varphi} & I_{\varphi\psi} \\
I_{\psi\varphi}    & I_{\psi\psi}    \\
\end{pmatrix}
\end{equation}

\begin{equation}
\gamma=
\begin{pmatrix}
\varphi \\ \psi
\end{pmatrix}^T \ \cdot
\begin{pmatrix}
I_{\varphi\varphi} & I_{\varphi\psi} \\
I_{\psi\varphi}    & I_{\psi\psi}
\end{pmatrix}  \cdot
\begin{pmatrix}
\varphi \\ \psi
\end{pmatrix}
\end{equation}

\noindent where $S$ is the previously defined covariance matrix. The min-max test aims to minimize the non-centrality parameter $\gamma$ with respect to the parameters that
are not suspected for change. The minimum of $\gamma$ with respect to $\psi$ is given for:

\begin{equation}
\psi^{*}=\arg\min\limits_{\psi}{\gamma}={\varphi^T}(I_{\varphi\varphi}-I_{\varphi\psi}
{I_{\psi\psi}^{-1}}I_{\psi\varphi})\varphi
\end{equation}

\noindent
and is found to be

\begin{equation}
\begin{tabular}{l}
$\gamma^{*}=\min\limits_{\psi}{\gamma} =
{\varphi}^T(I_{\varphi\varphi}-I_{\varphi\psi}{I_{\psi\psi}^{-1}} \
{I_{\psi\varphi}}){\varphi} =$\\
$=\begin{pmatrix}
\varphi  \\ -I_{\psi\psi}^{-1}I_{\psi\varphi}\varphi
\end{pmatrix}^T \
\begin{pmatrix}
I_{\varphi\varphi} & I_{\varphi\psi} \\
I_{\psi\varphi}    & I_{\psi\psi}
\end{pmatrix} \
\begin{pmatrix}
\varphi \\ {-I_{\psi\psi}^{-1}}{I_{\psi\varphi}}\varphi
\end{pmatrix}$
\end{tabular}
\end{equation}

\noindent
which results in

\begin{equation}
\begin{tabular}{l}
$\gamma^{*}=$\\
${\varphi^T} \{ [{\it{I},-I_{\varphi\psi}{I_{\psi\psi}^{-1}}}]{M^T}{\Sigma^{-1} \} \
\Sigma^{-1} \{ \Sigma^{-1}M [\it{I},-I_{\varphi\psi}{I_{\psi\psi}^{-1}}}] \}  \varphi$
\end{tabular}
\end{equation}

\noindent
The following linear transformation of the observations is considered :

\begin{equation}
X_{\phi}^{*}={[\it{I},-I_{\varphi\psi}{I_{\psi\psi}^{-1}}]}{M^T}{\Sigma^{-1}}X
\end{equation}

\noindent The transformed variable $X_{\phi}^{*}$ follows a Gaussian distribution $N(\mu_{\phi}^{*},I_{\phi}^{*})$
with mean:

\begin{equation}
{\mu_{\varphi}^{*}}={I_{\varphi}^{*}}\varphi
\end{equation}

\noindent
and with covariance :

\begin{equation}
{I_{\varphi}^{*}}=I_{\varphi\varphi}-I_{\varphi\psi}{I_{\psi\psi}^{-1}}I_{\psi\varphi}
\end{equation}

\noindent
The max-min test decides between the hypotheses :

\begin{eqnarray*}
& H_0^{*} : \mu^{*}=0 \\
& H_1^{*} : \mu^{*}=I_{\varphi}^{*}{\varphi}
\end{eqnarray*}

\noindent
and is described by :

\begin{equation} \label{chi2_test_min_max}
\tau_{\varphi}^{*}={{X_{\varphi}^{*}}^T}{{I_{\varphi}^{*}}^{-1}}{X_{\varphi}^{*}}
\end{equation}

\noindent The stages of fault detection and isolation (FDI), for the PDE system, with the use of the local statistical approach are summarized in the following table:

\begin{center}
\begin{tabular}{|l|}
\hline
\noindent Table I: Stages of the local statistical approach for FDI\\
\hline \\
1. Generate the residuals partial derivative given by Eq.(\ref{partial_derivative_residual}) \\
2. Calculate the Jacobian matrix J given by Eq.(\ref{Jacobian}) \\
3. Calculate the sensitivity matrix M given by Eq.(\ref{sensitivity_matrix}) \\
4. Calculate the covariance matrix S given by Eq.(\ref{covariance_matrix}) \\
5. Apply the $\chi^2$ test for change detection of Eq.(\ref{chi2_test}) \\
6. Apply the change isolation tests of Eq. (\ref{chi2_test_sensitivity}) or Eq.(\ref{chi2_test_min_max}) \\
\\
\hline
\end{tabular}
\end{center}

\section{Simulation tests}  \label{section 6 : Simulation_tests}

\subsection{Detection of faulty sensor nodes}

\noindent The proposed filtering scheme was tested in estimation and fault diagnosis for a wave equation of the form
of Eq. (\ref{nonlinear_wave_equation}) under unknown boundary conditions. Nonlinear 1D wave-type partial differential equations of this type appear in models of coupled oscillators. One can consider for example the forced damped sine-Gordon equation [\ref{SaaLev13}-\ref{GuoBil07}]

\begin{equation} \label{nonlinear_1D_wave_PDE}
\begin{tabular}{c}
${{\partial}{\phi} \over {{\partial}t^2}}+c{{{\partial}{\phi}} \over {{\partial}t}}-k{{\partial}^2{\phi} \over {{\partial}x^2}}+{\epsilon}sin(\phi)=l$
\end{tabular}
\end{equation}

\noindent where $c$, ${\epsilon}$ and $l$ are constants. This type of PDE appears in many physical phenomena, such as nonlinear resonant optics and Josephson junctions, or as a dynamic model of electrons in a crystal lattice. Eq. (\ref{nonlinear_1D_wave_PDE}) describes the motion of an array of pendula each of which is coupled to its nearest neighbors by a torsional spring with a coupling coefficient $k$. Each pendulum is subject to a constant torque $l$
and to a viscous drag force with coefficient $c$. The angle $x_i=\phi_i$ of the $i$-th pendulum and the vertical axis
evolves according to Eq. (\ref{nonlinear_1D_wave_PDE}).\\

\noindent To perform state estimation of the distributed parameter system of the Eq. (\ref{nonlinear_1D_wave_PDE}) a  grid consisting of $n=50$ points was considered. The number of measurement points was $n_1=25$ . In general, the number of measurements of the state variable $x_1=\phi(x,t)$ which can  be used by the estimation algorithm is  $n_1{\leq}n$, where $n$ is the number of grid points, and the criterion to select the number of grid points where measurements will be taken is to maintain the system's observability.  \\

%@+ \begin{figure} [htb]
%@+ \begin{center}
%@+ %$ \begin{array}{c@ {\hspace{0.2in}} c} \multicolumn{1}{l}{\mbox{\bf
%@+ % }} & \multicolumn{1}{l}{\mbox{\bf }} \\ [-0.53cm]
%@+ $\begin{tabular}{c}
%@+ \rotatebox{-90}{\epsfig{file=function_phi_v2.eps, width=60mm, height=60mm}} \\
%@+ \mbox{\bf (a)}
%@+ \\
%@+ \rotatebox{-90}{\epsfig{file=function_phi_der_v2.eps, width=60mm, height=60mm}} \\
%@+ \mbox{\bf (b)}
%@+ % \\ [0.4cm] \mbox{\bf (a)} & \mbox{\bf (b)}
%@+ %\end{array}$
%@+ \end{tabular}$
%@+ \end{center}
%@+ \caption{(a) monitored wave function $\phi(x,t)$ (b) first derivative of the wave function $\dot{\phi}(x,t)$}
%@+ \label{fig: DPS_function}
%@+ \end{figure}

\begin{figure}[htb]
\begin{center}
\rotatebox{-90}{\epsfig{file=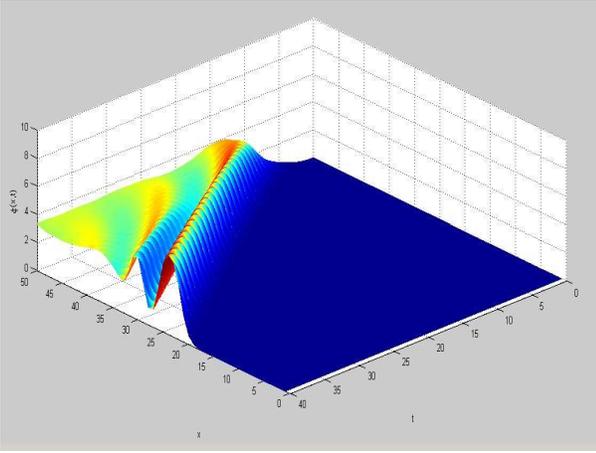, width=60mm,
height=80mm}} \caption{The monitored wave function $\phi(x,t)$} \label{fig: DPS_function}
\end{center}
\end{figure}

\begin{figure} [htb]
\begin{center}
% $\begin{array}{c@ {\hspace{0.05in}} c} \multicolumn{1}{l}{\mbox{\bf
% }} & \multicolumn{1}{l}{\mbox{\bf }} \\ [-0.53cm]
$\begin{tabular}{c}
{\epsfig{file=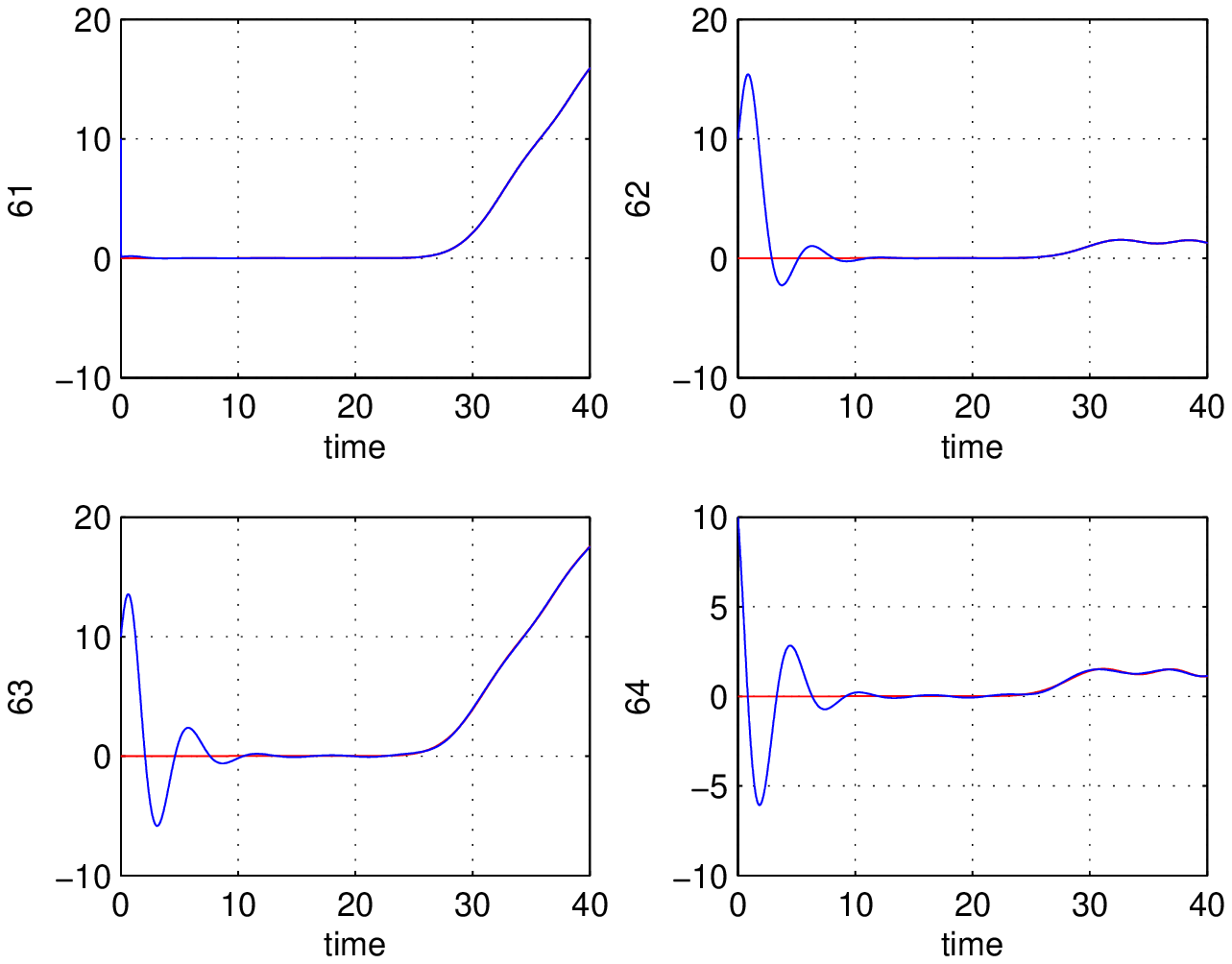, width=60mm, height=60mm}} \\
\mbox{\bf (a)} \\
{\epsfig{file=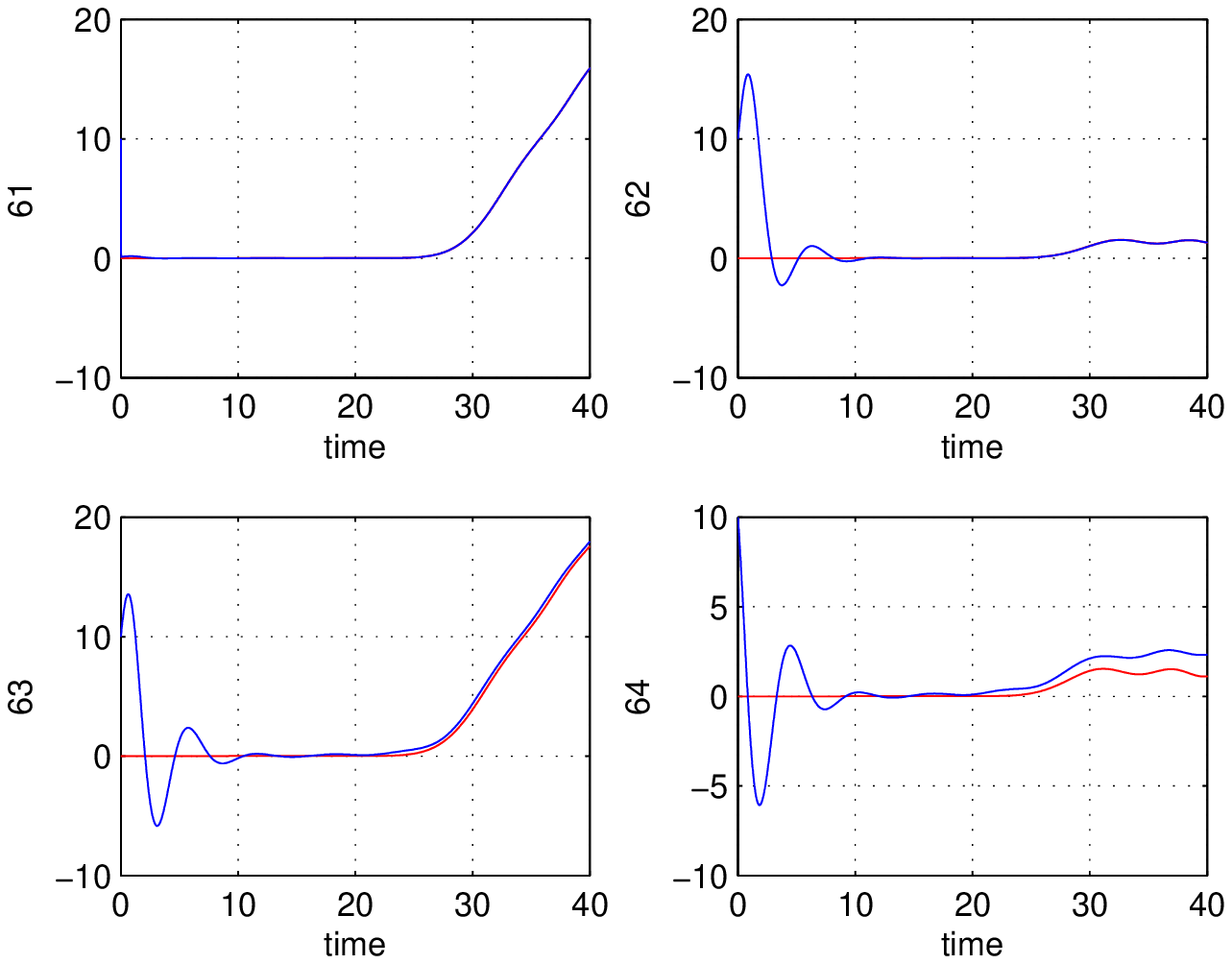, width=60mm, height=60mm}} \\
\mbox{\bf (b)}
% \\ [0.4cm] \mbox{\bf (a)} & \mbox{\bf (b)}
% \end{array}$
\end{tabular}$
\end{center}
\caption{Monitoring of state vector elements $y_{61}$ to $y_{64}$ of the DPS (a) when all sensors are fault-free (b)
when a fault appears at sensor monitoring output $22$, i.e. state variable $y_{85}$}
\label{fig: sensor_fault_y61_y64}
\end{figure}

\begin{figure} [htb]
\begin{center}
% $\begin{array}{c@ {\hspace{0.05in}} c} \multicolumn{1}{l}{\mbox{\bf
% }} & \multicolumn{1}{l}{\mbox{\bf }} \\ [-0.53cm]
$ \begin{tabular}{c}
{\epsfig{file=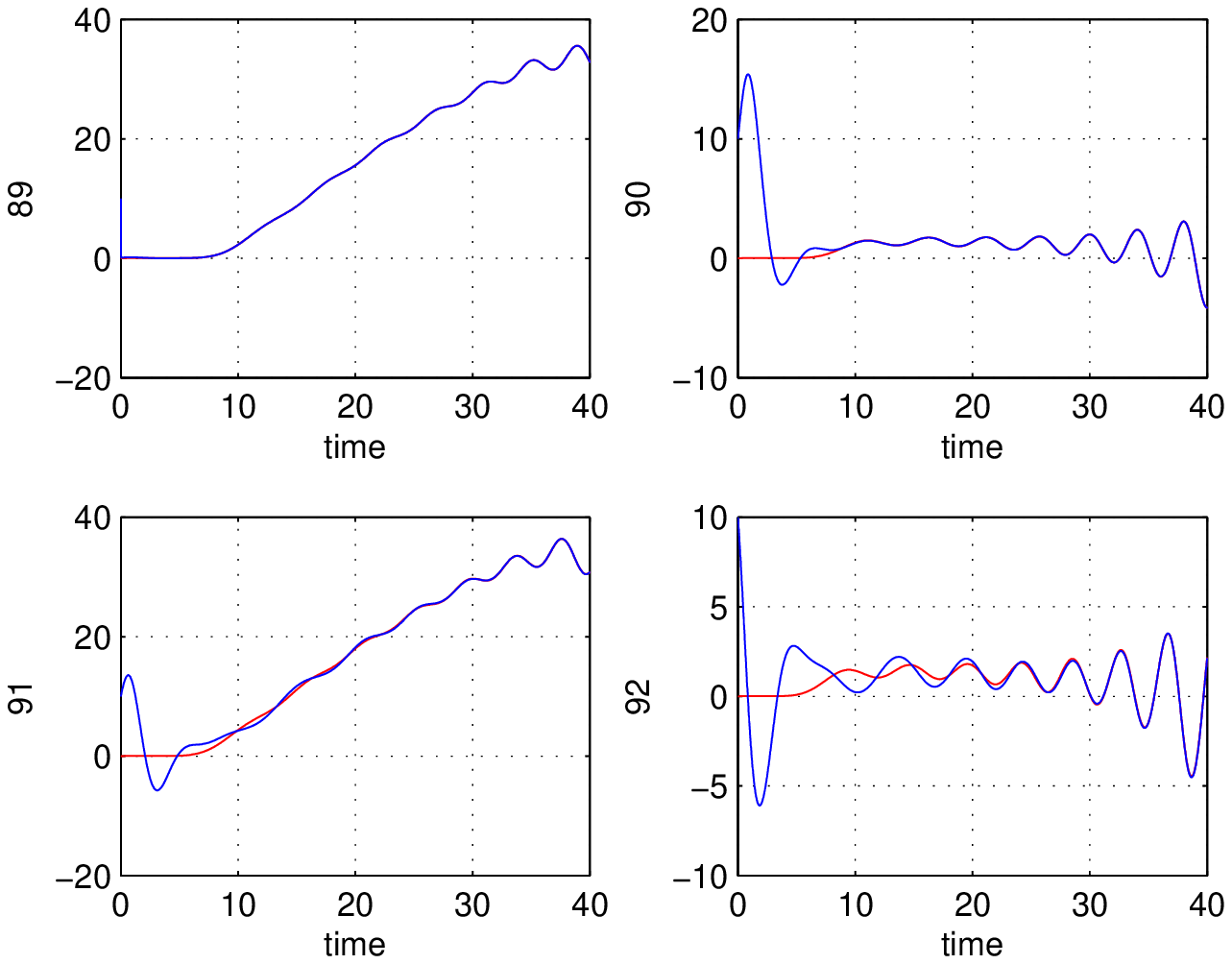, width=60mm, height=60mm}} \\
\mbox{\bf (a)} \\
{\epsfig{file=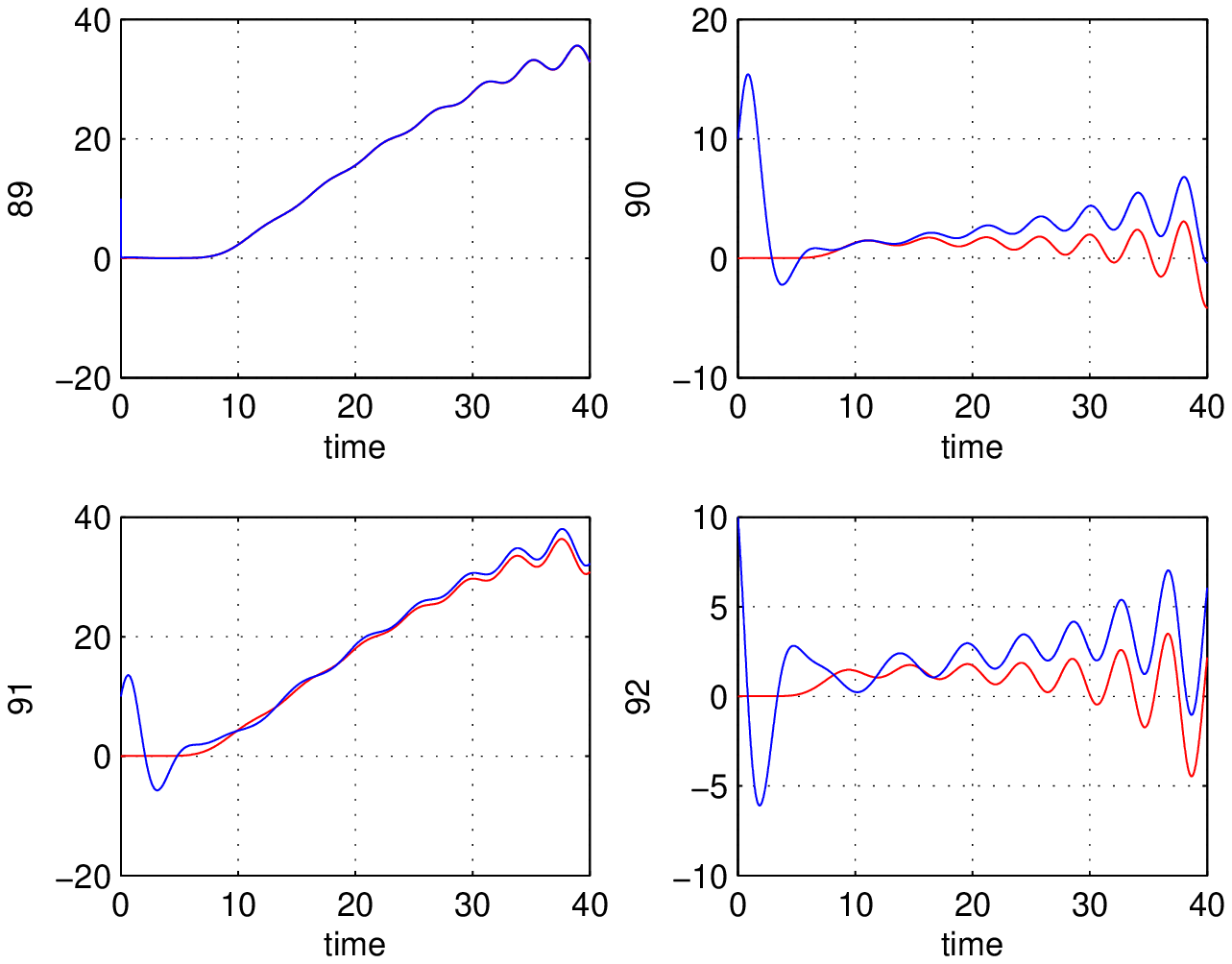, width=60mm, height=60mm}} \\
\mbox{\bf (b)}
% \\ [0.4cm] \mbox{\bf (a)} & \mbox{\bf (b)}
% \end{array}$
\end{tabular}$
\end{center}
\caption{Monitoring of state vector elements $y_{89}$ to $y_{92}$ of the DPS (a) when all sensors are fault-free (b)
when a fault appears at sensor monitoring output $22$, i.e. state variable $y_{85}$}
\label{fig: sensor_fault_y89_y92}
\end{figure}

\noindent In Fig. \ref{fig: DPS_function} the monitored wave function $\phi(x,t)$ and the associated first derivative in time $\dot{\phi}(x,t)$ is depicted. Indicative results about the estimation obtained at local grid points is given in Fig. \ref{fig: sensor_fault_y61_y64} and in Fig. \ref{fig: sensor_fault_y89_y92}. It was assumed that erroneous measurements were provided by one specific sensor used in the monitoring of the PDE system (e.g. sensor no 85 monitoring output 22 of the PDE system).  \\

\noindent The experimental results show that state estimates in the vicinity of the sensor subjected to fault (sensor monitoring output 22, i.e. state vector element 85) exhibit significant deviations comparing to the real values of the state vector. As moving far from the faulty sensor (e.g. state vector elements 61 to 64) the estimated and the real values of the state vector elements exhibit smaller differences. Thus, the comparison of the measurements recorded by the sensors against the estimated values from the Derivative-free nonlinear Kalman Filter provides a clear indication about the malfunctioning sensor. Thus is becomes possible to isolate this sensor from the rest of the sensors set. \\

\begin{figure} [htb]
\begin{center}
% $\begin{array}{c@ {\hspace{0.05in}} c} \multicolumn{1}{l}{\mbox{\bf
% }} & \multicolumn{1}{l}{\mbox{\bf }} \\ [-0.53cm]
$\begin{tabular}{c}
{\epsfig{file=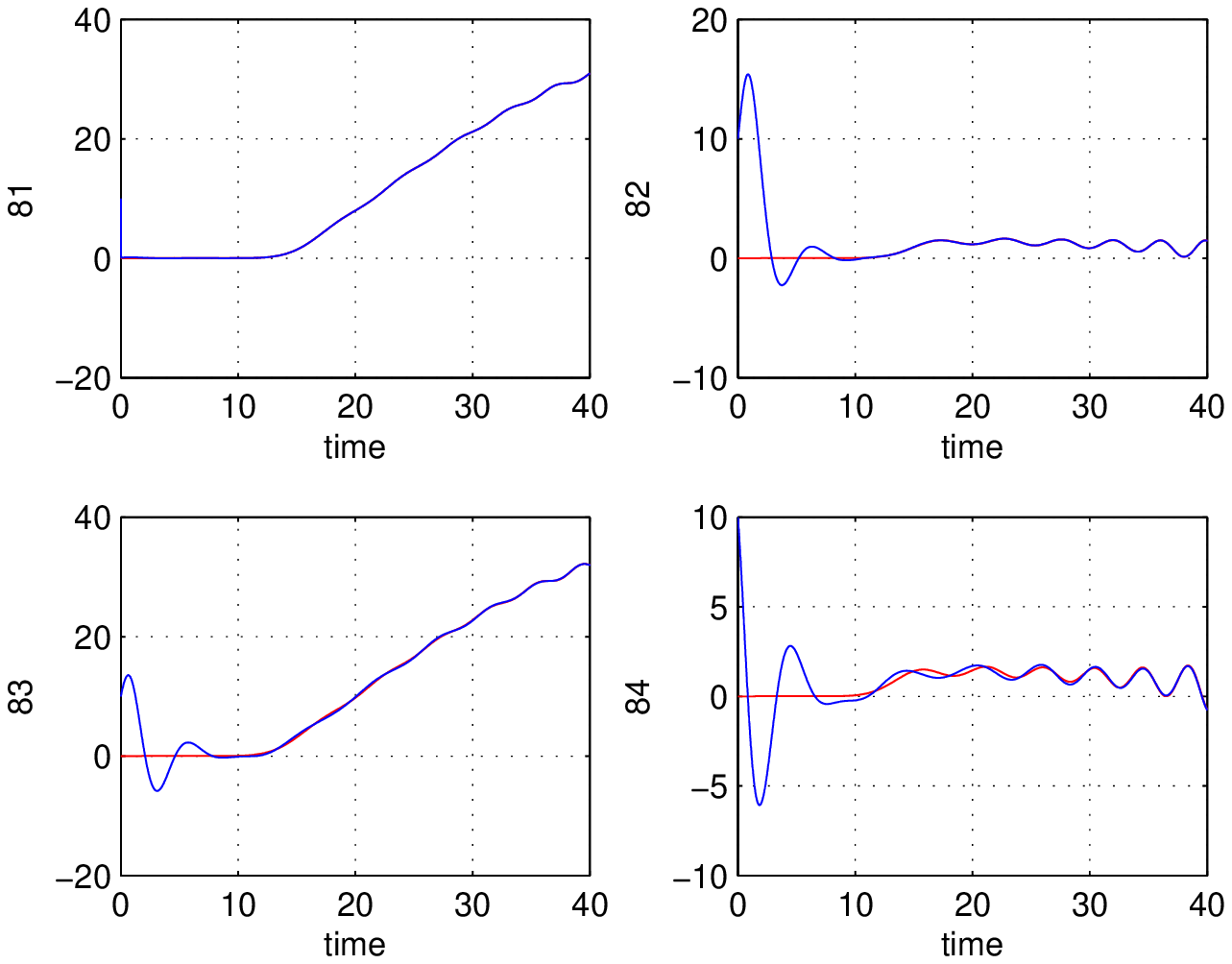, width=60mm, height=60mm}} \\
\mbox{\bf (a)} \\
{\epsfig{file=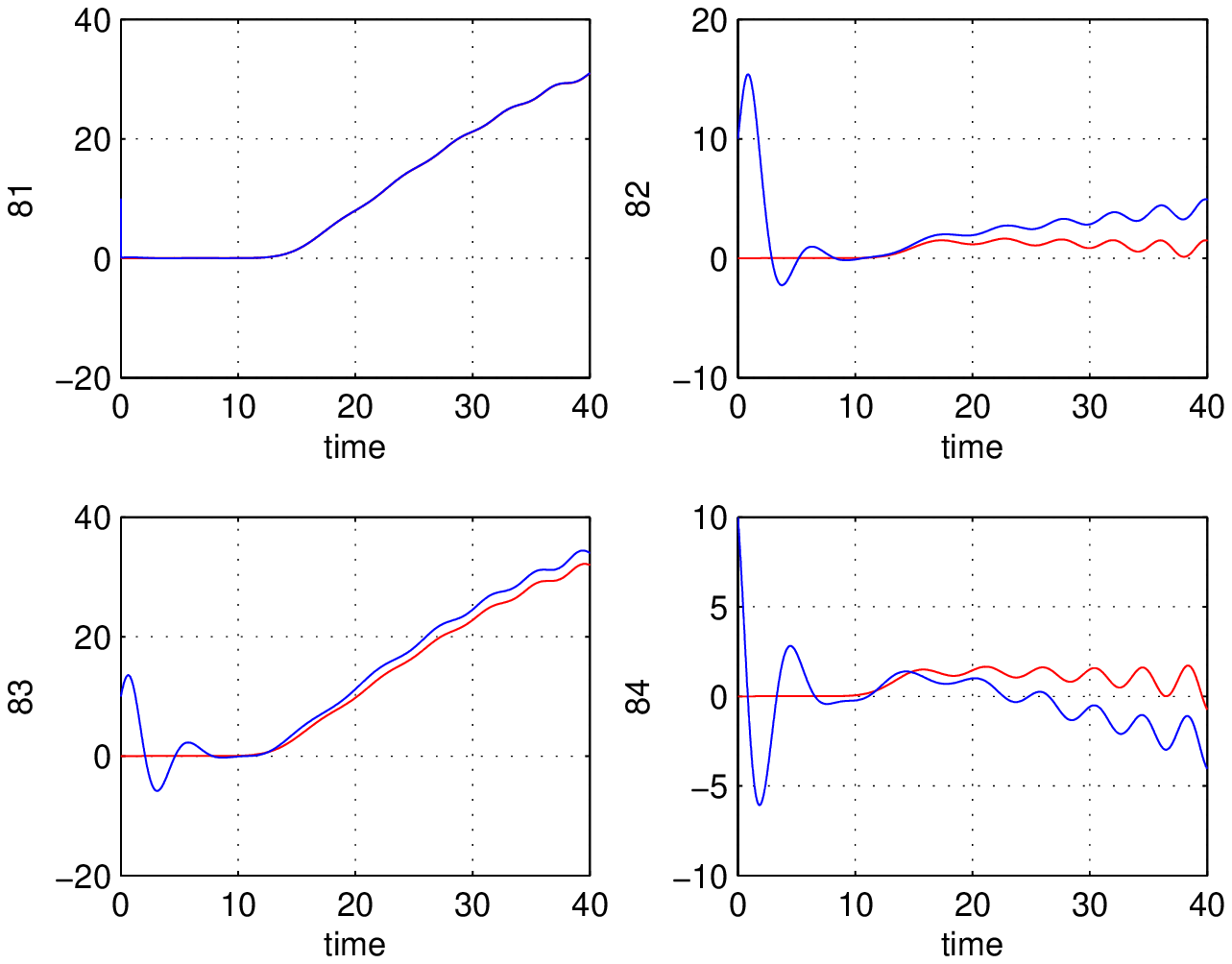, width=60mm, height=60mm}} \\
\mbox{\bf (b)}
% \\ [0.4cm] \mbox{\bf (a)} & \mbox{\bf (b)}
% \end{array}$
\end{tabular}$
\end{center}
\caption{Monitoring of state vector elements $y_{81}$ to $y_{84}$ of the DPS (a) when all sensors are fault-free (b)
when a fault appears at sensor monitoring output $22$, i.e. state variable $y_{85}$}
\label{fig: sensor_fault_y81_y84}
\end{figure}

\begin{figure} [htb]
\begin{center}
% $\begin{array}{c@ {\hspace{0.05in}} c} \multicolumn{1}{l}{\mbox{\bf
% }} & \multicolumn{1}{l}{\mbox{\bf }} \\ [-0.53cm]
$\begin{tabular}{c}
{\epsfig{file=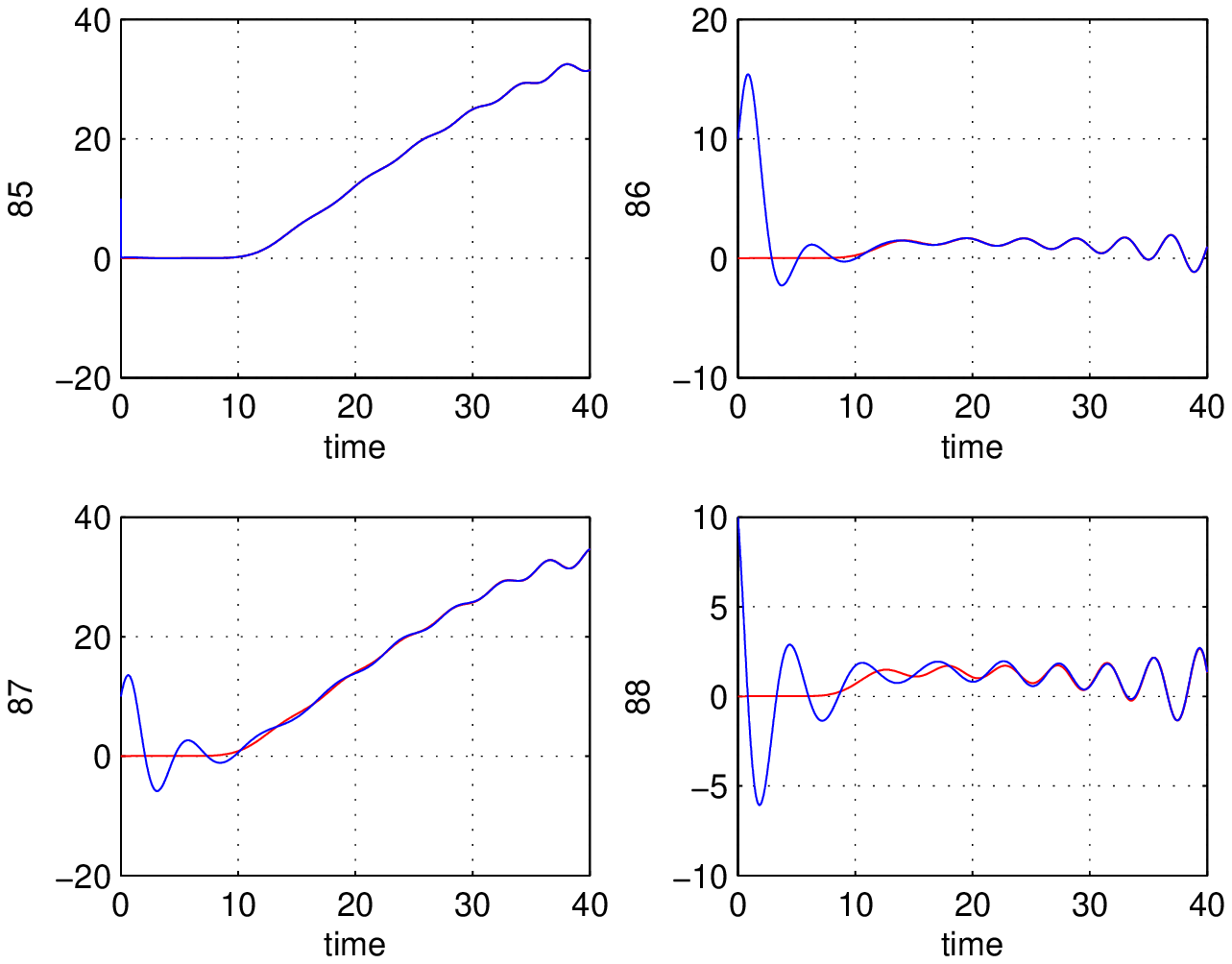, width=60mm, height=60mm}} \\
\mbox{\bf (a)} \\
{\epsfig{file=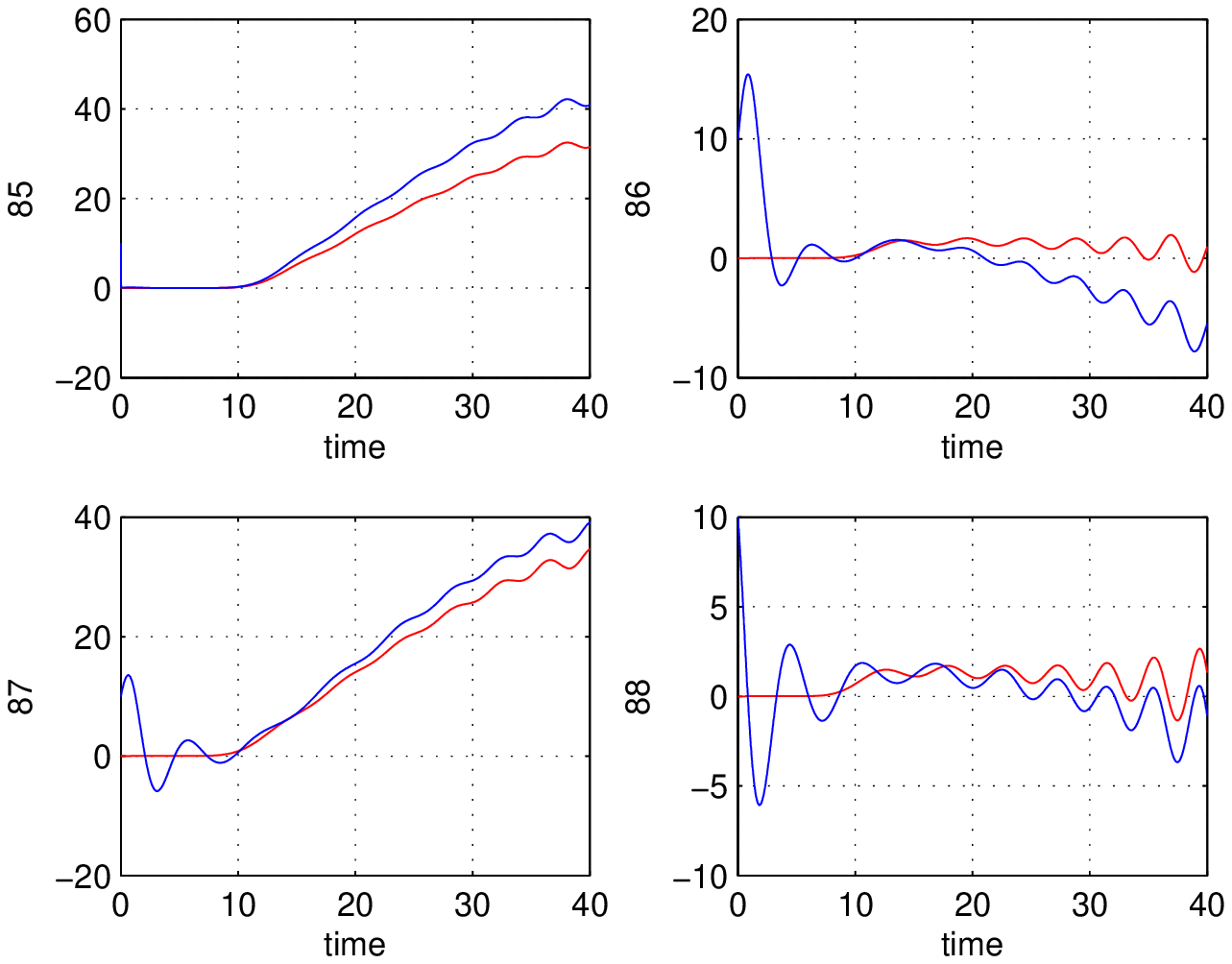, width=60mm, height=60mm}} \\
\mbox{\bf (b)}
% \\ [0.4cm] \mbox{\bf (a)} & \mbox{\bf (b)}
% \end{array}$
\end{tabular}$
\end{center}
\caption{Monitoring of state vector elements $y_{85}$ to $y_{88}$ of the DPS (a) when all sensors are fault-free (b)
when a fault appears at sensor monitoring output $22$, i.e. state variable $y_{85}$}
\label{fig: sensor_fault_y85_y88}
\end{figure}

%@+ \begin{figure} [p]
%@+ \begin{center}
%@+ $\begin{array}{c@ {\hspace{0.2in}} c} \multicolumn{1}{l}{\mbox{\bf
%@+ }} & \multicolumn{1}{l}{\mbox{\bf }} \\ [-0.53cm]
%@+ {\epsfig{file=y93_y96_fault_free_res1.eps, width=70mm, height=70mm}}
%@+ &
%@+ {\epsfig{file=y93_y96_fault_res1.eps, width=70mm, height=70mm}}
%@+ \\ [0.4cm] \mbox{\bf (a)} & \mbox{\bf (b)}
%@+ \end{array}$
%@+ \end{center}
%@+ \caption{Monitoring of state vector elements $y_{93}$ to $y_{96}$ of the DPS (a) when all sensors are fault-free (b)
%@+ when a fault appears at sensor monitoring output $22$, i.e. state variable $y_{85}$}
%@+ \label{fig: sensor_fault_y93_y96}
%@+ \end{figure}

%@+ \begin{figure} [p]
%@+ \begin{center}
%@+ $\begin{array}{c@ {\hspace{0.2in}} c} \multicolumn{1}{l}{\mbox{\bf
%@+ }} & \multicolumn{1}{l}{\mbox{\bf }} \\ [-0.53cm]
%@+ {\epsfig{file=y97_y100_fault_free_res1.eps, width=70mm, height=70mm}}
%@+ &
%@+ {\epsfig{file=y97_y100_fault_res1.eps, width=70mm, height=70mm}}
%@+ \\ [0.4cm] \mbox{\bf (a)} & \mbox{\bf (b)}
%@+ \end{array}$
%@+ \end{center}
%@+ \caption{Monitoring of state vector elements $y_{97}$ to $y_{100}$ of the DPS (a) when all sensors are fault-free
%@+ (b) when a fault appears at sensor monitoring output $22$, i.e. state variable $y_{85}$}
%@+ \label{fig: sensor_fault_y97_y100}
%@+ \end{figure}

\subsection{Change detection in the distributed parameter system\\}

\noindent It is also important to detect changes in the coefficients  of the distributed parameter model. To this end the the following state-space equation of the distributed parameter system is used. The state-space equation is written in the form of an ARMAX model which enables fault diagnosis with the use of the local statistical approach. \\

\noindent Without loss of generality it is assumed that coefficient $K$ remains the same
at all points of the grid. Using only the last subsystem in description of dynamics of the PDE one has

\begin{equation}
\begin{tabular}{c}
$\dot{y}_{1,N}=y_{2,N}$ \\
$\dot{y}_{2,N}={{K} \over {{\Delta}x^2}}\phi_{N+1}-{{2K} \over {{\Delta}x^2}}y_{1,N}+{{K} \over {{\Delta}x^2}}y_{1,N-1}+f(y_{1,N})$
\end{tabular}
\end{equation}

\noindent Defining $v_N={{K} \over {{\Delta}x^2}}\phi_{N+1}+{{K} \over {{\Delta}x^2}}y_{1,N-1}+f(y_{1,N})$ the following discrete-time description of the system is obtained

\begin{equation}  \label{discrete_state_space_model1}
\begin{tabular}{c}
$\begin{pmatrix}
y_{1,N}(k+1) \\
y_{2,N}(k+1)
\end{pmatrix}=
\begin{pmatrix}
1 & T_s \\
-{{2K} \over {Dx}^2}Ts & 1
\end{pmatrix}
\begin{pmatrix}
y_{1,N}(k) \\
y_{2,N}(k)
\end{pmatrix}+
\begin{pmatrix}
0 \\
T_s
\end{pmatrix}v_N(k)
$
\end{tabular}
\end{equation}

\begin{equation} \label{discrete_state_space_model2}
\begin{tabular}{c}
$z_{1,N}(k)=\begin{pmatrix}
1 & 0
\end{pmatrix}
\begin{pmatrix}
y_{1,N}(k) \\
y_{2,N}(k)
\end{pmatrix}$
\end{tabular}
\end{equation}

\noindent The associated ARMA model is found to be

\begin{equation} \label{ARMA_model1}
\begin{tabular}{c}
$z_{1,N}(k+2)=2z_{1,N}(k+1)-(1+{{{T_s^2}{2K}} \over {Dx^2}})z(k)+v_N(k)$
\end{tabular}
\end{equation}

\noindent In a similar manner, the discrete-time model of the Kalman Filter is written as

\begin{equation} \label{discrete_state_space_model1_KF}
\begin{tabular}{c}
$\begin{pmatrix}
\hat{y}_{1,N}(k+1) \\
\hat{y}_{2,N}(k+1)
\end{pmatrix}=
\begin{pmatrix}
1 & T_s \\
-{{2K} \over {Dx}^2}Ts & 1
\end{pmatrix}
\begin{pmatrix}
\hat{y}_{1,N}(k) \\
\hat{y}_{2,N}(k)
\end{pmatrix}+$\\
$+\begin{pmatrix}
0 \\
T_s
\end{pmatrix}v(k)+
\begin{pmatrix}
\kappa_1(k) \\
\kappa_2(k)
\end{pmatrix}\hat{e}(k)$
\end{tabular}
\end{equation}

\begin{equation} \label{discrete_state_space_model2_KF}
\begin{tabular}{c}
$\hat{z}(k)=\begin{pmatrix}
1 & 0
\end{pmatrix}
\begin{pmatrix}
\hat{y}_{1,N}(k)
\hat{y}_{2,N}(k)
\end{pmatrix}$
\end{tabular}
\end{equation}

\noindent and the associated ARMAX model is found to be

\begin{equation}\label{ARMA_model_KF}
\begin{tabular}{c}
$\hat{z}_{1,N}(k+2)=2\hat{z}_{1,N}(k+1)-(1+{T_s^2}{{2K} \over {Dx^2}})\hat{z}_{1,N}(k)+$\\
$+{T_s}v_N(k)+{\kappa_1(k)}\hat{e}(k+1)+(\kappa_1(k)-{\kappa_2(k)}T_s)\hat{e}(k)$
\end{tabular}
\end{equation}

\noindent where $\hat{e}(k)$ is the estimation error (innovation). The output of the ARMAX model can be also written in the product form

\begin{equation}\label{ARMA_model_KF_v2}
\begin{tabular}{c}
$\hat{z}_{1,N}(k+1)=w(k){\cdot}X^T(k)$
\end{tabular}
\end{equation}

\noindent where the weights vector is defines as

\begin{equation}\label{weights_vector}
\begin{tabular}{l}
$w(k)=$\\
$\begin{pmatrix}
2 & -(1+{T_s^2}{{2K} \over {Dx^2}}) & {T_s} & {\kappa_1(k)} & (\kappa_1(k)-{\kappa_2(k)}T_s)
\end{pmatrix}$
\end{tabular}
\end{equation}

\noindent and the regressor vector is defined as

\begin{equation} \label{regressor_vector}
\begin{tabular}{l}
$X(k)=$\\
$\begin{pmatrix}
\hat{z}_{1,N}(k) &  \hat{z}_{1,N}(k-1) & v_N(k-1) & \hat{e}(k) & \hat{e}(k-1)
\end{pmatrix}$
\end{tabular}
\end{equation}

\begin{figure} [htb]
\begin{center}
% $\begin{array}{c@ {\hspace{0.2in}} c} \multicolumn{1}{l}{\mbox{\bf
% }} & \multicolumn{1}{l}{\mbox{\bf }} \\ [-0.53cm]
$\begin{tabular}{c}
{\epsfig{file=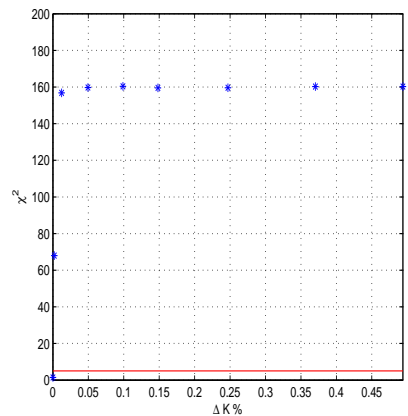, width=60mm, height=60mm}} \\
\mbox{\bf (a)} \\
{\epsfig{file=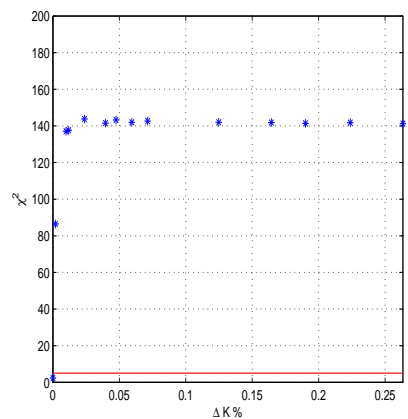, width=60mm, height=60mm}} \\
\mbox{\bf (b)}
% $ \\ [0.4cm] \mbox{\bf (a)} & \mbox{\bf (b)}
% \end{array}$
\end{tabular}$
\end{center}
\caption{Detection of incipient changes in the coefficient $K$ of the nonlinear wave equation using the $\chi^2$ test (a) nominal value $K=0.04050$  (b)  nominal value $K=0.05050$}
\label{fig: global_chi2_test}
\end{figure}

\noindent Indicative results about the performance of the global $\chi^2$ in detection of incipient changes of parameter $K$ of the nonlinear wave PDE are depicted in Fig. \ref{fig: global_chi2_test}. The fault threshold is set equal to the number of parameters in the associated ARMAX model, i.e. $\eta=5$. It can be observed that the proposed FDI method was capable of detecting changes in parameter $K$ which were less than $1\%$ of the coefficient's nominal value. For small deviations from the parameter's nominal value the $\chi^2$ test obtained a value that was several times larger than the fault threshold.

\section{Conclusions} \label{Section 7: Conclusions}

\noindent The paper has proposed state estimation and fault diagnosis in distributed parameter systems with a new nonlinear filtering approach, the so-called Derivative-free nonlinear Kalman Filter. The method is based into decomposition of the nonlinear Partial Differential equation that describes the dynamics of the distributed parameter system, into a set of nonlinear ordinary differential equations. Next, with the application of a change of coordinates (diffeomorphism) that is based on differential flatness theory, the local nonlinear differential equations are turned into linear ones. This enables to describe the PDE dynamics with a state-space equation that is in the linear canonical (Brunovsky) form. For the linearized equivalent of the PDE system it is possible to perform state estimation with the use of the standard Kalman Filter recursion. Unlike other nonlinear filtering methods, such as the Extended Kalman Filter, the proposed approach does not require the computation of partial derivatives and Jacobian matrices. Moreover, it avoids the cumulative numerical errors which appear in distributed Extended Kalman Filtering and which are due to truncation of higher order terms in the Taylor expansion of the system's dynamical model. Thus, the proposed filtering method succeeds improved accuracy in the estimates of the dynamics of the distributed parameters system.

\noindent Next, the \textit{local statistical approach to fault diagnosis} has been proposed for performing fault diagnosis of distributed parameter systems. To this end, statistical processing of the differences (residuals) between the Kalman filter's outputs and measurements from the distributed parameters system have undergone statistical processing. Fault diagnosis with the Local Statistical Approach has two significant advantages: (i) it provides a credible criterion ($\chi^2$ test) to detect changes in the parameters of the PDE system. This criterion is more efficient than the normalized square error and mean error tests since it employs the modeling error derivative and records the tendency for change. Thus early change detection for distributed parameters system becomes possible (ii) it recognizes the parameters of the PDE system that have undergone a change. Thus fault isolation becomes possible as well. The efficiency of the Derivative free nonlinear Kalman Filter in fault diagnosis  has been confirmed through simulation experiments in the case of distributed parameter systems described by 1D-wave equations.

%%%%%%%%%%%%%%%%%%%%%%%%%

\end{document}